\documentstyle[preprint,aps,eqsecnum]{revtex}

\begin{document}
\draft
 \widetext
\newcommand{\Sp}{\,}

\title{Exclusonic Quasiparticles and Thermodynamics of \\
Fractional Quantum Hall Liquids}
\author{Y.S. Wu$^1$, Y. Yu$^2$,
Y. Hatsugai$^3$ and M. Kohmoto$^4$}
\address{$^1$Department of Physics, University of Utah,
Salt Lake City, UT 84112, U.S.A.}
\address{$^2$Institute of Theoretical Physics, Academia
Sinica, Beijing 100080, China}
 \address{$^3$Department of Applied Physics,
University of Tokyo, Bunkyo-ku, Tokyo 113, Japan}
\address{$^{4}$Institute for Solid State Physics,
University of Tokyo, Roppongi, Minato-ku,
Tokyo 106, Japan}

\maketitle
\widetext
\begin{abstract}
\leftskip 54.8pt
\rightskip 54.8pt

Quasielectrons and quasiholes in the fractional
quantum Hall liquids obey fractional (including
nontrivial mutual) exclusion statistics. Their
statistics matrix can be determined from several
possible state-counting scheme, involving different
assumptions on statistical correlations. Thermal
activation of quasiparticle pairs and thermodynamic
properties of the fractional quantum Hall liquids
near fillings $1/m$ ($m$ odd) at low temperature
are studied in the approximation of generalized
ideal gas. The existence of hierarchical states
in the fractional quantum Hall effect is shown
to be a manifestation of the exclusonic nature of
the relevant quasiparticles. For magnetic
properties, a paramagnetism-diamagnetism
transition appears to be possible at finite
temperature.
\end{abstract}

\vspace{.1in}

\section{Introduction}

It is well-known that quantum statistics of a particle
(or elementary excitation) plays a fundamental
role in determining statistical or thermodynamic
properties of a quantum many-body system.
Bose-Einstein and Fermi-Dirac statistics are two well
established ones, which are central to many familiar
or novel phenomena involving many particles.
For example, superfluidity or superconductivity
is essentially due to Bose-Einstein condensation;
and the stability of macroscopic matter is known
to depend crucially on the Fermi-Dirac statistics
of electrons. Since early days of
quantum mechanics, an outstanding problem has
been to search for a {\it generalization} of or even an
{\it interpolation} between these two statistics.
Mathematically, of course, there exist many possibilities.
But as physicists we are interested in what are
{\it physically relevant}, in the sense that the new
statistics must be realized in physical systems existing
in nature (or at least, in models which describe some
interesting aspects of real physics).

In the last two decades or so,
the interest in this search has become stronger
and stronger in the study of lower dimensional condensed
matter systems. For quite a while, it has been recognized
that situations which interpolate between bosons and
fermions may appear in one- \cite{1,Suther} and
two-dimensional \cite{4} many-body systems, though no
discussion of relevant statistical distributions
until recently.

By now there have been (at least) two distinct
ways to define fractional statistics:

\noindent 1) by examining the change in phase of
a multi-particle wave function due to the exchange of
two identical particles,

\noindent  2) by counting the number of
independent multi-particle quantum states to formulate a
generalization of Pauli exclusion principle.

\noindent For the usual quantum statistics, i.e. for
bosons and fermions, the above two methods are
equivalent to each other, in spite of their conceptual
difference. For fractional statistics, however, they
are generally inequivalent and we distinguish
between the two definitions by calling them as
``exchange statistics'' and ``{\it exclusion statistics}''
respectively, and the corresponding particles as
``anyons''  and ``{\it exclusons}'' . Fractional exchange
statistics or anyons \cite{4} have been first explored
in the study of  quasiparticles in two (space) dimensional
systems, such as fractional quantum Hall (FQH) liquids
and anyon superconductivity.

Recently Haldane \cite{2} has formulated, by counting
many-body states, a generalized Pauli exclusion principle
in arbitrary spatial dimensions. Based on this idea, one
of us (YSW) \cite{3} have defined generalized ideal gas
for particles obeying such fractional (including mutual)
exclusion statistics, and have formulated its quantum
statistical mechanics and thermodynamics. These new
definitions are not merely mathematical construction;
they have been shown to be realized as the exotic
statistics obeyed by elementary excitations in certain
one, two or higher dimensional strongly correlated
systems \cite{2}-\cite{HKKW}. In this paper, we discuss
an important case: quasiparticles in the fractional
quantum Hall (FQH) effect.

By now it is well-known that the ground state of the
two-dimensional electron gas in a strong perpendicular
magnetic field with electron filling factor $\nu=1/m$
($m$ odd integer) is an incompressible quantum liquid
\cite{Laughlin}, and that it has two species  of
quasiparticle excitations, both of which are
fractionally charged  \cite{Laughlin} ($e^{*}_{\pm}
=\mp 1/m$) anyons \cite{Halperin,Arovas}. However, the
anyon approach is not very suitable for calculating
low-temperature thermodynamic properties of the FQH
liquids, since thermal activation of quasiparticle
pairs is directly governed by the counting law for
many-body states rather than the law for exchange phases.
This is where the concept of exclusion statistics comes
into play. The FQH quasiparticles are known to be
strongly correlated. The key issue here is to clarify
how strong correlation between quasiparticles manifests
itself in the state-counting. Is possible not only
fractional exclusion of single-particle states for
identical quasiparticles, but also mutual exclusion
of states between quasihole and quasielectron, which
cannot be dealt with in the anyon approach.

The many-body state counting for FQH
quasiparticles is a subtle problem, whose study
started with Haldane's paper in 1983 \cite{Hald1}.
Now we still do not have a full answer yet, except
for the low-lying excited states of the FQH liquids,
which however should be enough to account for
low-temperature thermodynamic properties.
There are several possible assignments for
the statistics matrix of FQH quasiparticles,
involving different assumptions concerning the
nature of the quasiparticles and their statistical
correlations. (See below, Sec. III, for details.)
Fortunately, these assignments of statistics matrix
can be tested by numerical simulation on small
systems. In addition, one would also like
to put these assignments to experimental tests,
which needs comparing theoretical predictions
with experimental data. As a first step towards
this, we have calculated thermodynamic properties
of the FQH liquids at low temperatures, based on the
dilute gas approximation for thermally activated
quasiparticles: When the quasiparticles are dilute,
we may ignore Coulomb interactions between them, and
apply the statistical thermodynamic formalism for
generalized ideal gas given in ref. \cite{3}, which
incorporates mutual statistics between different
species of quasiparticles. It is hoped that this
approximation could be improved in the future by
including the effects of Coulomb interactions between
quasiparticles. Our thermodynamic calculation is
done with three different assignments of statistics
matrix, with the hope that one day the experimental
measurements of thermodynamic properties of FQH
liquids might distinguish between them,
providing information about the statistical
correlations between FQH quasiparticles.

This paper is organized as follows. We first
review in Sec. II the state counting definitions of
exclusion statistics, including mutual statistics
between non-identical (quasi)particles. Then in Sec.
III it is shown that quasielectrons and quasiholes
in the FQH liquids with fillings $1/m$ ($m$ odd) obey
such fractional exclusion statistics, and the
determination of their statistics matrix from various
physical arguments or working hypotheses is reviewed,
with several somewhat different outcomes. Using the
statistical distribution \cite{3} for generalized
ideal gas, we study in Sec. IV and V, with analytic
and numerical methods, thermal activation of
quasiparticle pairs near filling $1/m$ ($m$ odd).
Then we show in Sec. VI that the occurrence
of new (hierarchical) incompressible states,
corresponding to divergent pressure, at appropriate
fillings at $T=0$ is a manifestation of fractional
exclusion statistics, while at finite temperature
the pressure of the system can never become divergent.
In Sec. VII we compute low-temperature thermodynamic
properties of FQH liquids, including magnetic
properties. The final section VIII is devoted to
conclusions and discussions.

\section{Exclusion Statistics}

The definition of fractional exclusion statistics is
directly based on state-counting, a basic concept in
quantum statistical mechanics. It is well-known that
bosons and fermions have different counting for
many-body states, or different statistical weight $W$:
The number of quantum states of $N$ identical
particles occupying a group of $G$ states, for
bosons or fermions respectively, is given by

\begin{equation}
W_{b}= {(G+N-1)! \over N!~ (G-1)!}~,~~~
{\rm{or}}~~ W_{f}= {G! \over N!~ (G-N)!}~.    \label{1}
\end{equation}

\noindent A simple generalization and interpolation is

\begin{equation}
 W = {[G+(N-1)(1-\alpha)]!
\over N!~ [G-\alpha N-(1-\alpha)]!}~,         \label{2}
\end{equation}

\noindent with $\alpha=0$ corresponding to
bosons and $\alpha=1$ fermions.
The physical meaning of this equation is
the following: By assumption, the statistical weight
remains to be {\it a single combinatoric number}, so one
can count the states by thinking of the particles
{\it effectively as bosons}, with
the effective number of available single-particle
states being {\it linearly dependent on the
particle number}:

\begin{equation}
G_{eff}^{(b)} = G - \alpha (N-1).
\label{3}
\end{equation}

\noindent Obviously, for genuine bosons ,
$G_{eff}^{(b)}\;$ is independent of the
particle number. In all other cases,
$G_{eff}^{(b)}$ is linearly
dependent on the particle number.
This is the defining feature of
the fractional exclusion statistics.
The statistics parameter $\alpha$ tells us,
on the average, how many single-particle
states that a particle can exclude others
to occupy. For $\alpha \neq 1$, this
generalizes the Pauli exclusion principle
for one species.

It is easy to generalize this state
counting to more than one species:

\begin{equation}
W = {\prod}_i ~ { [G_i + N_{i}-1 -
\sum_j \alpha_{ij}(N_j-\delta_{ij})]!
\over (N_i)!~ [G_i - 1
- \sum_j \alpha_{ij}(N_j-\delta_{ij})]! }~.   \label{4}
\end{equation}

\noindent Here $G_i$ is the number of states
when the system consists of only a single particle.
By definition, the diagonal $\alpha_{ii}$ is
the ``self-exclusion'' statistics of species $i$,
while the non-diagonal $\alpha_{ij}$ (for $i\neq j$)
is the mutual-exclusion statistics. Note that
$\alpha_{ij}$, which Haldane \cite{2} called {\it
statistical interactions}, may be
{\it asymmetric} in $i$ and $j$. The
interpretation is similar to the one species
case. The number of available single-particle
states for species $i$, in the presence of
other particles, is again linearly dependent on
particle numbers of all species:

\begin{equation}
G^{(b)}_{eff,i}= G_i -\sum_{j}
\alpha_{ij} (N_j - \delta_{ji}).
\label{Geff}
\end{equation}

We note that as a generalized Pauli exclusion principle,
eq. (\ref{2}) or (\ref{4}) implies strong correlations
between the particles, and does not give rise to the
same state counting as the old generalization suggested
in ref. \cite{Gentile}, in which particles independently
fill a fixed number of single-particle states with the
constraint that at most $n$ particles are allowed in
one and the same state.

Some remarks on exclusion statistics are in order:

1) This definition of exclusion statistics
is {\it independent of spatial dimensionality}
of the system, in contrast to the exchange
statistics of anyons which has a connection to the
braid group, making sense only in two spatial
dimensions \cite{Wu1}.

2) In contrast to anyons, there is {\it no periodicity}
in exclusion statistics parameter $\alpha$, so it
makes sense to consider the cases with $\alpha>1$ or
even $\alpha>2$.

3) The state-counting definition of exclusion
statistics naturally allows {\it mutual statistics}
from the beginning, implying that exclusion may occur
between states of different species, a completely new
situation we have not been faced before in statistical
mechanics.

\section {Exclusion Statistics for
FQH Quasiparticles}

There are two kinds of quasiparticles in the
Laughlin $1/m$-liquid: quasiholes labeled by $-$ and
quasiparticles labeled by $+$. In this paper we treat
them as two distinct species and demonstrate that
their many-body states obey the counting law given by
(\ref{Geff}), with an appropriate $2\times 2$ statistics matrix
$\alpha_{ij}$ with $i,j=+,-$.  The statistics matrix depends
on the nature of FQH quasiparticles and their correlations.
Several scenarios are possible in this regard. In this
section,  we are going to discuss four possible
scenarios for FQH quasiparticles that have appeared
in the literature: 1) anyons in the lowest Landau
level,  2) bosonic vortices, 3) composite fermions,
4) correlated vortices or composite fermions. They
differ in the assumption of whether certain correlations,
such as hard-core constraints, exist between the
quasiparticles or not, leading to subtle difference
in statistics matrix. It is remarkable that the
statistics matrix can be subject to numerical test
for small systems on a sphere. We are not going to talk
about the details, but will briefly summarize the
status of such numerical tests and quote relevant
references when appropriate.

\subsection {Anyons in the lowest Landau level}

In determining exclusion statistics of FQH
quasiparticles, let us first try to explore the
fact that they are fractionally charged anyons.
(Though later we will see that the picture of
non-interacting anyons is not very suitable for
calculating thermal activation of quasiparticle pairs.)

Good trial electron wave functions for states
with quasiparticles in the $1/m$ FQH liquid were
first proposed by Laughlin \cite{Laughlin}.
In these wave functions the coordinates of the
quasiparticles appear as parameters (or collective
coordinates). If one moves very slowly the
coordinates of one quasiparticle, say a quasihole,
around a closed loop in the FQH liquid, the
electron wave function acquires a Berry
phase, which can be interpreted as the phase
due to the motion of the quasiparticle
traveling along the loop. As shown in ref.
\cite{Arovas}, the Berry phase is always
proportional to the number of electrons
enclosed in the loop. If the loop encloses
none of other quasiparticles, the Berry phase
is the same as that for a charge in a magnetic
field, with electrons acting as quantized
sources of ``flux''. Thus, a quasiparticle sees
the electrons just like an electron sees
the external magnetic field. When the loop
encloses another quasiparticle, say a quasihole,
the change in the Berry phase is due to a deficit in
the number of enclosed electrons caused by
the enclosed quasihole, and it is attributed
to the exchange phase of the two quasiholes,
showing that they are anyons with fractional
exchange statistics \cite{Arovas}. Combining
the two results, one is led to a simple picture
that the FQH quasiparticles are anyons in
the lowest Landau level of a fictitious
magnetic field, whose strength is determined
by the density of electrons. Indeed, the wave
function for the FQH quasiparticles suggested
by Halperin \cite{Halperin.hier} are such as if the
quasiparticles are in the lowest Landau level.


Now let us count the states of  $N$
non-interacting anyons (of one species)
in the lowest Landau level. Though not for
all levels and all states, a number of exact
solutions for anyons in a magnetic field
have been known \cite{14}. Among them,
fortunately, are the complete set of solutions
for all anyons in the lowest Landau level
(if the number of anyons is less than the
Landau degeneracy). The total (ground state)
energy turns out to be the sum of the cyclotron
energy of individual particles, independent
of the exchange statistics of anyons:

\begin{equation}
E = N \varepsilon_{c}/2~.
\label{energytot}
\end{equation}


To count the states, we consider anyons in
a circular disk with a fixed size.  In the symmetric
gauge, besides the usual Gaussian factor, the many-anyon
wave function is known to be of the form (with $z_{i}$
the complex coordinates of electrons):

\begin{equation}
    \Psi = \prod_{i< j} (z_{i}-z_{j})^{\theta/\pi}
    \cdot \Phi (z_{1}, \ldots{}, z_{N}),
\label{Jastrow}
\end{equation}

\noindent with anyon statistics $0\leq\theta <2\pi$. But
now in the lowest Landau level, the function $\Phi$ is a
symmetric polynomial of $(z_{1}, \ldots{}, z_{N})$.
The state counting can be easily done by looking at
the symmetric polynomial $\Phi$ (counting as bosons).
However, the fixed-size condition requires a fixed
highest angular momentum, or a fixed highest
power of a single variable $z_{i}$ in the wave
function $\Psi$. On the other hand, the Jastrow-type
prefactor $\Pi_{i<j}(z_i-z_j)^{\theta/\pi}$,
implying non-vanishing relative angular momenta
between anyons, takes away some powers of $z_{i}$
and reduces the degree of the polynomial $\Phi$.
Alternatively, an anyon can see the statistical flux
of other anyons, which in the present case is opposite
to the external magnetic flux. Therefore, in the boson
counting, {\it with size and external flux fixed},
the effective Landau degeneracy is determined by
the external magnetic flux $N_{\phi}$ less the
anyon statistical flux $(\theta/\pi) (N-1)$: \cite{16}

\begin{equation}
G_{eff}^{(b)}= \frac{N_{\phi}}{m}
-\frac {\theta}{\pi} (N-1).
\label{anyboson}
\end{equation}

\noindent Hence  eq. (\ref{2}) applies, with
the single anyon Landau degeneracy
$G=N_{\phi}/m$, and the exclusion statistics
for anyons in the lowest Landau level can be
read off from eq. (\ref{3}):

\begin{equation}
\alpha =\theta /\pi. 
\label{anyexclu}
\end{equation}

In Ref. \cite{Arovas}, the exchange statistics has
been shown to be $\theta_{-} =\pi/m$ for quasiholes,
and $\theta_{+} =-\pi/m$ for quasielectrons.
Thus,  eq. (\ref{anyexclu}) leads to the following
diagonal exclusion statistics for quasiparticles:

\begin{equation}
\alpha_{--} = \frac{1}{m},\quad\quad
\alpha_{++}= \Biggl\{ {\begin{array}{ll}
  -1/m & ({\rm soft-core}), \\
  2-1/m & ({\rm hard-core}).
 \end{array}}
\label{hardcore}
\end{equation}

\noindent  For the case of quasielectron, we note that
there are two possibilities: Since conceptually exchange
statistics is an angular parameter, defined only up to
a period of $2\pi$, $\theta =-\pi/m$ is equivalent to
$\theta = (2-1/m)\pi $. However, exclusion statistics is
always unambiguously defined and non-periodic at all;
i.e. the exclusion effects with $\alpha=-1/m$ and with
$\alpha=2-1/m$ are very different. So when one wants
to apply eq. (\ref{anyexclu}), he or she has to choose
between the possible two values of $\theta$. One may
notice that the wave function (\ref{Jastrow}) with
$\theta=-\pi/m$ is singular at $z_i=z_j$. Based on the
braid group, one may argue that the many-anyon wave
function (\ref{1}) should vanish as two anyons approach
each other, thus preferring $\theta/\pi=2-1/m$ over
$-1/m$ in the Jastrow-type prefactor and therefore
$\alpha=2-1/m$ over $-1/m$ for the exclusion statistics
for quasielectrons. Obviously, the former value of
$\alpha$ leads to stronger exclusion between
quasielectrons, as if they have "hard-core".

Whether the quasielectrons really satisfy the "hard-core
constraint" can be tested by numerical experiments. Such
numerical experiments have been done by three groups,
\cite{16}, \cite{Su} and \cite{20}, for electrons with
Coulomb interactions on a sphere in the field of
a monopole at its center. Their results unambiguously
support the exclusion statistics (\ref{hardcore}) with the
"hard-core" value for quasielectrons.

\subsection{Bosonic Vortex Scheme}

The above scenario for FQH quasiparticles as anyons
in the lowest landau level has the disadvantage that it
tells us nothing about the mutual (or non-diagonal)
statistics between quasihole and quasielectron, which
is important for studying thermal activation of
quasiparticle pairs. So we need  other, more direct
ways to count states for quasiparticles in the Laughlin
$1/m$-liquid.

A fundamental relation which will play a key role
in state-counting is the ``total-flux'' constraint

\begin{equation}
N_{\phi}\equiv eBV/hc = mN_{e} + N_{-} - N_{+}~,
\label{flux-constraint}
\end{equation}

\noindent between the electron numbers $N_{e}$
and quasiparticle numbers $N_{-}$ and $N_{+}$. The
basic idea behind this relation is the following
observation \cite{Laughlin}: To generate a quasiparticle
in the incompressible Laughlin liquid, one may pierce
the droplet by an infinitely thin solenoid and
slowly turn on magnetic flux inside it. When the flux
reaches a flux quantum, a quasiparticle will be formed
around the solenoid; whether it is a quasihole or
quasielectron depends on the direction of the solenoid
flux (parallel or anti-parallel to the external
magnetic field).

Further state counting relies on the assumptions on
statistical correlations of quasiparticles. There are
different counting schemes based on the bosonic
vortex picture, the composite fermion picture and
variation of both. Let us consider them in turn.

The bosonic vortex scheme is based on the picture
that the FQHE quasiparticles are vortex-like
excitations in the incompressible planar quantum liquid,
and for a fixed number of excitations we count their
states as if they are bosons. Assuming only the
minimal (quantized) circulation, there are two
possible orientations for vortex circulation on the
plane, corresponding to quasihole and quasielectron
respectively. So what is essential to this counting
scheme is to determine the number of available states
for each species of vortices.

This can be inferred from an observation by Haldane
and Wu \cite{19} that for vortices in a planar
quantum liquid, their core $X$- and
$Y$-coordinates do not commute with each
other, as if they were the guiding-center
coordinates for a charged particle in
a magnetic field, with fluid particles (i.e. electrons
in the present case) as sources of quantized flux.
Thus, the number of available states for vortex-like
excitations is essentially determined by the ``Landau
degeneracy'' of this fictitious magnetic field, or
the number of electrons:

\begin{equation}
G_{eff,-}^{(b)}=N_{e}, \;\;\;\;\;
G_{eff,+}^{(b)}=N_{e}.
\label{boson1}
\end{equation}

To derive the exclusion statistics for FQH
quasiparticles, one needs to fix the external
flux $N_{\phi}$. So let us express $G_{eff,\mp}$
in terms of $N_{\phi}$ by eliminating $N_{e}$
from these equations with the help of the constraint
(\ref{flux-constraint}). Then we obtain

\begin{eqnarray}
G_{eff,-}^{(b)} &=& \frac{1}{m} N_{\phi} -\frac{1}{m} N_{-}
+ \frac{1}{m} N_{+}\, , \nonumber \\
G_{eff,+}^{(b)} &=& \frac{1}{m} N_{\phi} -\frac{1}{m} N_{-}
-(- \frac{1}{m}) N_{+} \, .
\label{Haldscheme}
\end{eqnarray}

The first term on the right side gives the single-quasiparticle
degeneracy in terms of the external flux:

\begin{equation}
G_{+}=G_{-}= (1/m)N_{\phi},
\label{Gpm}
\end{equation}
so the proportionality constant $1/m$ is identified as
the fractional charge (absolute value) of the quasiparticles.
And the coefficients of $N_{\mp}$ give the
statistics matrix:

\begin{eqnarray}
\begin{array}{cclccl}
\alpha_{++} & =& -1/m, & \alpha_{+-}&=& 1/m,  \\
\alpha_{-+} & =& -1/m, & \alpha_{--} &=&1/m~.
\end{array}
\label{alpha-vortex1}
\end{eqnarray}
\noindent This result was first derived by Haldane \cite{2}.

Comparing with eq. (\ref{hardcore}), we note that
$\alpha_{++}=-1/m$ here corresponds to
soft-core quasielectrons. For hard-core quasielectrons,
eq. (\ref{boson1}) should be replaced by

\begin{equation}
G_{eff,-}^{(b)}=N_{e}, \;\;\;\;\;
G_{eff,+}^{(b)}=N_{e} - 2 (N_{+}-1),
\label{boson2}
\end{equation}

\noindent with the second term in $G_{eff,+}^{(b)}$
representing the exclusion effects due to the
{\it hard core} of quasielectrons. In the presence
of an external magnetic field, the two orientations
of vortex circulation are not equivalent, so there is
an asymmetry between quasiholes and quasielectrons.
Physically, the hard-core nature of quasielectrons
may be due to electron number surplus in the core of
quasielectrons. This leads to \cite{comm1}

\begin{eqnarray}
G_{eff,-}^{(b)} &=& \frac{1}{m} N_{\phi} -\frac{1}{m} N_{-}
+ \frac{1}{m} N_{+}\, , \nonumber \\
G_{eff,+}^{(b)} &=& \frac{1}{m} N_{\phi} -\frac{1}{m} N_{-}
-(2- \frac{1}{m}) N_{+} \, ,
\label{Haldscheme2}
\end{eqnarray}
resulting in a statistics matrix somewhat different from
eq. (\ref{alpha-vortex1}):

\begin{eqnarray}
\begin{array}{cclccl}
\alpha_{++} & =& 2-1/m, & \alpha_{+-}&=& 1/m,  \\
\alpha_{-+} & =& -1/m, & \alpha_{--} &=&1/m~.
\end{array}
\label{alpha-vortex2}
\end{eqnarray}

We note that now the diagonal statistics $\alpha_{++}$
in eq.  (\ref{alpha-vortex2}) agree with that of
hard-core quasielectrons in eq.  ({\ref{hardcore}).
However, here it has been possible to demonstrate
nontrivial mutual statistics between quasihole and
quasielectron. This means that with $N_{\phi}$ fixed,
the presence of quasielectrons will affect the number of
``available'' states for quasiholes and {\it vice versa}.

\subsection{Composite Fermion Scheme}

The central idea of the composite fermion approach
is that the FQH state of electrons in a physical magnetic
field can be explained as the IQH state of composite
fermions in an effective magnetic field \cite{Jain}.
Imagine an adiabatic process in which we somehow
collect $2p$ ($p$ an integer) flux quanta to each
electron to form an electron-flux composite.  The
additional Aharonov-Bohm phase, due to the attached
flux, associated with moving one composite around
another is $e ^{i 2p\pi} =1$. So the statistics of the
composite remains to be the same as the electron, motivating
the name of composite fermion. These composites are now
moving in a reduced magnetic field  $B_{eff}=B-2 \pi (2p \rho) $,
where $\rho$ is the density of electrons, which is the same as
the density of composite fermions. (Recall that in our convention,
the unit of flux is $2 \pi$.) The filling factor for composite
fermions then increases to $\nu _{eff}$, given by
$\nu ^{-1}_{eff}=(B-4\pi p\rho)/2\pi \rho=\nu^{-1}-2p$. For
$\nu _{eff}=n$ ($n$ an integer), we have

\begin{equation}
\nu=\frac {n}{2pn+1}.
\label{filling}
\end{equation}
Thus, fractional Hall systems with $\nu = n/(2pn+1)$
may be adiabatically changed into an integer Hall system
with filling factor $n$, as was also emphasized by Greiter
and Wilczek \cite{Greiter}. Note that this argument
gives us more than we had hoped for. The case we
wanted to understand, with $\nu^{-1}$= odd,
is obtained for $n=1$.

Let us do state counting for the state with $n=1$,
or $\nu=1/(2p+1)$.
The Landau degeneracy for the composite fermion in the
residue magnetic field  is given by the effective flux

\begin{equation}
N_{\phi,eff}= N_{\phi} - 2p N_{e}\, ,
\label{Neff1}
\end{equation}

\noindent while the number of excitations are determined by

\begin{equation}
N_{\phi,eff}= N_{e} + N_{-} - N_{+}\, ,
\label{Neff2}
\end{equation}

\noindent Eliminating $N_{\phi,eff}$ from these two
equations, we recover the same constraint
(\ref{flux-constraint}) as before with $m=2p+1$.

The number of available single-particle states for
unit-charged composite-fermion excitations is obviously

\begin{equation}
G_{eff,\mp}=N_{\phi,eff} - (N_{\mp}-1) \, .
\label{compfermi}
\end{equation}

\noindent Here we have assumed that the magnetic
field is so strong that we can ignore the possibility
for quasielectron to fill Landau levels higher than
the lowest available one.

To derive the true charge and statistics of the
quasiparticle excitations, we need to express
$N_{\phi,eff}$ in  eq. (\ref{compfermi})
in terms of the external $N_{\phi}$,
resulting in \cite{comm1}

\begin{eqnarray}
G_{eff,-}^{(b)} &=& \frac{1}{m} N_{\phi} -\frac{1}{m} N_{-}
-(1- \frac{1}{m}) N_{+}\, , \nonumber \\
G_{eff,+}^{(b)} &=& \frac{1}{m} N_{\phi} +(1-\frac{1}{m}) N_{-}
-(2- \frac{1}{m}) N_{+} \, .
\label{Jainscheme}
\end{eqnarray}

\noindent Here we have used eqs. (\ref{Neff1}) and
(\ref{Neff2}). The coefficient of $N_{\phi}$ recovers
the fractional charge (absolute value) $1/m$ for the
quasiparticles; see eq. (\ref{Gpm}). From the
coefficient of the other terms one reads off the
exclusion statistics:

\begin{eqnarray}
\begin{array}{cclccl}
\alpha_{++} & =& 2-1/m, & \alpha_{+-}&=& -1+1/m,  \\
\alpha_{-+} & =& 1-1/m, & \alpha_{--} &=&1/m~.
\end{array}
\label{jainstat}
\end{eqnarray}

\subsection{Correlated Vortex or Projected
Composite Fermion Scheme}

Comparing eq. (\ref{jainstat}) with eq. (\ref{alpha-vortex2}),
we see that the composite fermion scheme leads to the
same diagonal statistics both for quasiholes and for
quasielectrons as  the bosonic vortex scheme, with
quasielectrons being automatically {\it hard-core}.
Thus it is not surprising that the two schemes give
the same prediction about the occurrence of hierarchical
states at $T=0$, since only one species of quasiparticles
exist at $T=0$ when the filling factor deviates from
the magic $1/m$, so that only diagonal statistics
is relevant.

However, mutual statistics in eq. (\ref{jainstat}) and eq.
(\ref{alpha-vortex2}) obtained from the above two
schemes are obviously different. Which is correct?
Or neither is correct? To decide, one needs
to study situations in which both species
of quasiparticles coexist at the same time.
This problem has been numerically studied in ref.
\cite{SuWuYang} (see also \cite{Isakov2}).
It turns out that neither of the mutual
statistics given in eq. (\ref{alpha-vortex2})
and eq. (\ref{jainstat}) is correct. The correct
statistics matrix, for low-lying excitations,
turns out to be

\begin{eqnarray}
\begin{array}{cclccl}
\alpha_{++} & =& 2-1/m, & \alpha_{+-}&=& -2+1/m,  \\
\alpha_{-+} & =& 2-1/m, & \alpha_{--} &=&1/m~.
\end{array}
\label{jainstat2}
\end{eqnarray}

In the bosonic vortex scheme, this can be obtained by
incorporating certain amount of mutual exclusion (or
inclusion) between vortices and anti-vortices in
eq. (\ref{boson2}) as follows:

\begin{eqnarray}
G_{eff,-}^{(b)} &=& N_{e} - 2N_{+}\; ,\nonumber \\
G_{eff,+}^{(b)} &=&N_{e} + 2N_{-} - 2 (N_{+}-1)\; .
\label{boson3}
\end{eqnarray}
We call this scheme as the correlated vortex scheme, since
in ref. \cite{SuWuYang} it has been shown that this
modification is due to the necessity of inserting  some
"hard-core" Jastrow factor between quasihole and
quasielectron in the quasiparticle wave functions, which
represents correlations of a new type between the vortex
and the anti-vortex.

To reproduce the statistics (\ref{jainstat2}) in the
composite fermion scheme, one needs to modify
eq. (\ref{compfermi}) to  \cite{Wu,Isakov2}

\begin{eqnarray}
G_{eff,-}&=& N_{\phi,eff} - (N_{-}-1) - N_{+}\; ,\nonumber \\
G_{eff,+}&=& N_{\phi,eff} + N_{-}- (N_{+}-1)\; .
\label{compfermi2}
\end{eqnarray}
The mutual exclusion added in these equations between
composite fermionic holes and composite fermionic
electrons can be interpreted as a consequence of the
{\it projected} composite fermion scheme: In the
composite fermion transformation quasielectron
states involve wave functions in the second Landau
level, so it is necessary to project the wave functions
down to the lowest Landau level to obtain the correct
many-electron wave functions. (For details, see
ref. \cite{WuJain}.) Indeed, the state counting resulting
from the above equation has been checked \cite{Wu,Isakov2}
to be indeed in agreement with the numerical data given in
ref. \cite{WuJain}, which verifies the necessity for
the projected composite fermion scheme.

We note that in either scheme, the mutual
statistics between quasihole and quasiparticle
are {\it anti-symmetric} rather than symmetric.

In summary, the bosonic vortex scheme (\ref{boson2})
and the unprojected composite fermion scheme
(\ref{compfermi}) lead to different mutual statistics
(\ref{alpha-vortex2}) and (\ref{jainstat}) respectively.
But the correlated vortex scheme (\ref{boson3})
and the projected composite fermion scheme
(\ref{compfermi2}) lead to the same statistics
matrix (\ref{jainstat2}). Numerical data favor the
latter. But we feel that there is no harm to leave
 these possibilities open to experimental tests. In the
following, we calculate thermodynamic properties of
FQH liquids with the three statistics matrices, with
the hope that someday experiments might be able to
distinguish between them.

\section {Thermal Activation of  FQH Quasiparticle Pairs}

It is well-known that the low-temperature
thermodynamics of a many-body system is
determined by its low-lying excited states above
the ground state. For FQH liquids at hand, our
fundamental assumption is that  their
{\it low-lying} excited states are dominated
by {\it weakly coupled} quasiparticles. Thus
their low-temperature thermodynamic properties
are determined by {\it thermal activation of FQH
quasiparticle pairs}. At low temperatures, when
the activated pairs are not very dense, one
may ignore their interaction energies.
Then the densities $\rho_{\pm}$ of the
excitations should be determined by the laws
for generalized ideal gas \cite{3} with two species,
in which the following two conditions are satisfied:

\noindent
1) The state-counting (\ref{4}) for statistical
weight $W$ is applicable.

\noindent
2) The total energy (eigenvalue)
is always of the form of a simple sum, in which
the $i$-th term is linear in the particle number
$N_{i}$:

\begin{equation}
 E=\sum_{i} N_{i} \varepsilon_{i},
\label{6}
\end{equation}
\noindent
with $\varepsilon_{i}$ identified as the energy
of a quasihole ($i=-$) or a quasielectron ($i=+$).
Though this condition (\ref{6}) is very natural
for {\it weakly} interacting FQH quasiparticles,
we note it is not compatible with non-interacting
anyons, except for {\it only one species}
of anyons all in the lowest Landau level
(see eq. (\ref{energytot})). This problem does
not exist in the theoretical framework of
exclusion statistics: The condition (\ref{6}) is
compatible with free exclusons, as exemplified
\cite{5}-\cite{HKKW} in one-dimensional
exactly solvable many-body models such as
the $\delta$-function repulsive
boson gas \cite{1} and the Calegero-Sutherland
model \cite{Suther}. This is one of the main
theoretical advantages of exclusion statistics
over exchange (or anyon) statistics in dealing
with statistical mechanics. (Moreover, the anyon
picture can not deal with more than one species,
so it is not suitable for studying thermal activation,
which involves both quasielectrons and quasiholes
and is expected to be a good place to look for
the effects of mutual statistics, with increasing
density of activated pairs.)

With theses assumptions, now we
are able to derive quantum statistical
mechanics of FQH quasiparticles.
Consider a grand canonical ensemble
at temperature $T$ and with chemical
potential $\mu_{i}$ for species $i=+,-$,
whose partition function is given by
(with $k$ the Boltzmann constant)

\begin{equation}
Z= \sum_{\{N_{i}\}} W(\{N_{i}\})~
\exp \{\sum_{i=+,-} N_{i} (\mu_{i} -\varepsilon_{i})/kT \}~.
\label{8}
\end{equation}

\noindent As usual, we expect that for very large
$N_{i}$, the summand has a
very sharp peak around the set of most-probable
(or mean) particle numbers $\{\bar{N}_{i}\}$.
Using the Stirling formula and
introducing the average ``occupation number
per state'' defined by
$n_{i} \equiv \bar{N}_{i}/ G_{i}$,
{}from the maximum condition

\begin{equation}
{\partial \over \partial n_{i}}\,
\bigl[ \log W + \sum_{j=+,-}
G_{j} n_{j}\,(\mu_{j} -
\varepsilon_{j})/kT \bigr] =0~,               \label{9}
\end{equation}
one obtains the equations determining the
most-probable distribution of $n_{i}$

\begin{equation}
\sum_{j=+,-} (\delta_{ij}w_j(T) +\alpha_{ij}) n_j(T) = 1~,
\label{n(T)}
\end{equation}
with $w_i(T)$ being determined by the functional
equations

\begin{eqnarray}
{w_{+}}^{\alpha_{++}} (1+w_{+})^{1-\alpha_{++}}
\Bigl( \frac {w_{-}}{1+w_{-}} \Bigr)^{\alpha_{-+}}
& = &  e^{(\varepsilon_{+}-\mu_{+})/kT }~, \nonumber \\
{w_{-}}^{\alpha_{--}} (1+w_{-})^{1-\alpha_{--}}
\Bigl( \frac{w_{+}}{1+w_{+}} \Bigr)^{\alpha_{+-}}
& = &  e^{(\varepsilon_{-}-\mu_{-})/kT }~.             \label{wpm}
\end{eqnarray}
{}From eqs. (\ref {n(T)}) and (\ref {wpm}),
$n_{\pm}$ are expressed in terms of  $w_{\pm}$ as

\begin{eqnarray}
n_{+}(T) & = & \frac {\rho_{+}(T)}{ \rho_{0}}  =  \frac {
w_{-}+\alpha_{--}-\alpha_{+-}} {(w_{+}+\alpha_{++}) (w_{-}+\alpha_{--})
-\alpha_{+-}\alpha_{-+} }~, \nonumber \\
n_{-}(T) & = & \frac{\rho_{-}(T)}{\rho_{0}}  =  \frac{
w_{+}+\alpha_{++}-\alpha_{-+}} {(w_{+}+\alpha_{++}) (w_{-}+\alpha_{--})
-\alpha_{+-}\alpha_{-+}}~,
\label{npm}
\end{eqnarray}
where $\rho_{0}\equiv G_{\pm}/V$, and
$\rho_{+}(T)$ and $\rho_{-}(T)$ are the density of
quasielectrons and quasiholes respectively.
The ratio $R(T)$ of the numbers of
quasielectrons and quasiholes is given by

\begin{equation}
R(T)\equiv {n_{+}(T) \over n_{-}(T)} =
{ w_{-}+\alpha_{--}-\alpha_{+-}
\over  w_{+}+\alpha_{++}-\alpha_{-+} }\, .
\label{R(T)}
\end{equation}

According to charge conservation, only
quasielectron-quasihole pairs are thermally
activated,  since they have opposite charges.
Thus, $N_{+}-N_{-}$ is independent
of temperature. Then the total-flux constraint
(\ref{flux-constraint}) implies that

\begin{equation}
n_+(T)-n_-(T)=m \delta,
\label{n+-n-}
\end{equation}
where $\delta=m(\nu-\nu_0),~(\nu_0=1/m)$. Thus
$\delta/m$ gives the deviation of the filling from $1/m$. Then
from eq. (\ref{npm}) one has

\begin{equation}
m\delta = { w_{-}-w_{+}+\alpha_{--}-
\alpha_{+-}+\alpha_{-+}-\alpha_{++}
\over (w_{+}+\alpha_{++}) (w_{-}+\alpha_{--})
-\alpha_{+-}\alpha_{-+} }~.
\label{mdelta}
\end{equation}
Charge conservation also requires that the chemical
potentials for the two species should satisfy $\mu_{+}+\mu_{-}=0$.
Multiplying the two equations in (\ref{wpm}), and
using the above constraints, one can derive a polynomial equation

\begin{equation}
{w_+}^{\alpha_{++}+\alpha_{+-}}{w_-}^{\alpha_{-+}+\alpha_{--}}
(1+w_{+})^{1-\alpha_{++}-\alpha_{+-}}
(1+w_{-})^{1-\alpha_{-+}-\alpha_{--}} =
e^{(\varepsilon_{+}+\varepsilon_{-})/kT }.             \label{wkt}
\end{equation}

Once $w_+$ and $w_-$ are determined from eqs. (\ref{mdelta})
and (\ref{wkt}), the $T$-dependent densities of both species
$\rho_{+}=n_+ \rho_0$ and $\rho_{-}=n_- \rho_0$ are given
by eq. (\ref{npm}). The mutual-statistics-dependent
thermodynamic potential $\Omega=-kT\log Z$  and  entropy $S$ are

\begin{equation}
\Omega \equiv - PV = -kT \sum_{i=+,-} G_i
\log {\rho_0+\rho_i - \sum_{j=+,-} \alpha_{ij} \rho_j
\over \rho_0 - \sum_{j=+,-} \alpha_{ij} \rho_j}~;
\label{Omega}
\end{equation}

\begin{equation}
{S\over k}= \sum_{i=+,-} G_i
\Bigl\{ n_i {\varepsilon_i - \mu_{i} \over kT} +
\log {\rho_0+\rho_i - \sum_{j=+,-} \alpha_{ij} \rho_j
\over \rho_0 - \sum_{j=+,-} \alpha_{ij} \rho_j} \Bigr\}.
\label{Entropy}
\end{equation}
Further the total entropy is written as
$S=\sum_{i}N_{i}s_{i}$ with

\begin{eqnarray}
{s_{i}\over k} &=&
 [1 + {\rho_{0}\over \rho_{i}} -
\sum_j \alpha_{ij} {\rho_{j}\over \rho_{i}}]\;
\log [1+\sum_j (\delta_{ij}-\alpha_{ij}) {\rho_{j}\over \rho_{0}}]
\nonumber\\
&&- \log {\rho_{i}\over \rho_{0}}
- ({\rho_{0}\over \rho_{i}}
- \sum_j \alpha_{ij} {\rho_{j}\over \rho_{i}})\;
\log (1 -\sum_j \alpha_{ij} {\rho_{j}\over \rho_{0}}).
\label{62}
\end{eqnarray}
\noindent Other thermodynamic functions, such as
specific heat and magnetization per unit area, follow
straightforwardly. For example, magnetization
per unit area is given by
\begin{equation}
{\cal M}
= \sum_{i} \Bigl( -\mu_{i} \rho_{i}
+ {e T \over m hc}
\log { \rho_{0} + \rho_{i}
-\sum_{j} \alpha_{ij} \rho_{j}
\over \rho_0 -\sum_{j} \alpha_{ij} \rho_{j}}\Bigr).
\label{61}
\end{equation}

\noindent Here $\mu_{\pm}=\partial \varepsilon_{\pm}
/\partial B$; and we have assumed the independence of
$\mu_{\pm}$. Hopefully, when  $T$ is of order of
$\varepsilon_{\pm}$ or higher, the
$\alpha_{ij}$-dependent second term may give
an appreciable contribution. Note that the
thermodynamic properties at the two sides of
electron filling $\nu_0\equiv N_{e}/N_{\phi}=1/m$
are not symmetric, due to asymmetry of quasielectron and
quasihole in self-exclusion and mutual-exclusion statistics.

\section{Explicit solutions near $\nu=1/m$}

We present some explicit formulas for thermal activation
of FQH quasiparticle pairs for three statistics matrices
(\ref{alpha-vortex2}), (\ref{jainstat}) and (\ref{jainstat2}),
which were derived from three different counting
schemes in Sec. III.

\subsection{Bosonic Vortex Picture}

{}From the statistics matrix (\ref{alpha-vortex2}),
eq. (\ref{npm}) determines the occupation number
of quasielectrons and of quasiholes to be

\begin{equation}
n_{+}={w_{-}\over (w_{+}+2-1/m) (w_{-}+1/m) +1/m^2 }~,
\label{v.n_+}
\end{equation}
and

\begin{equation}
n_{-}= {w_{+}+2\over (w_{+}+2-1/m) (w_{-}+1/m) +1/m^2 }~.
\label{v.n_-}
\end{equation}
where $w_+$ and $w_-$ can be obtained by solving
eqs. (\ref{mdelta}) and (\ref{wkt}), which are explicitly

\begin{equation}
m \delta w_+ w_{-} +(\delta+1)w_{+}
+[(2m-1)\delta-1]w_{-}+2+2\delta =0;
\label{v.mdelta}
\end{equation}
\begin{equation}
w_+^2(1+w_+)^{-1}(1+w_-)= e^{\Delta/kT}\equiv f(T),
\label{v.f(T)}
\end{equation}
where $\Delta =\varepsilon_{+}+\varepsilon_{-}$ is the pair
excitation gap.

The last two equations can not be solved
analytically, but numerical solution is possible. See
Figures \ref{fig-1-nplus} (a) and \ref{fig-2-nminus} (a)
for a two-dimensional plot of  $n_+(T,\delta)$ and
$n_-(T,\delta)$, obtained  numerically
for near $1/m=1/3$.

\subsection{Composite fermion scheme }

In this scheme, the statistics matrix is given by
eq. (\ref{jainstat}). Then eq. (\ref{npm}) is explicitly

\begin{equation}
n_{+}=
{ w_{-}+1
\over (w_{+}+2-1/m) (w_{-}+1/m) +(1-1/m)^2 }~,
\label{cf.n_+}
\end{equation}
and
\begin{equation}
n_{-}= { w_{+}+1
\over (w_{+}+2-1/m) (w_{-}+1/m) +(1-1/m)^2 }~.
\label{cf.n_-}
\end{equation}
with $w_+$ and $w_-$ determined by
\begin{equation}
w_+w_-=e^{\Delta/kT}\equiv f(T),
\label{f(T)}
\end{equation}
\begin{equation}
m\delta w_+w_{-} +(\delta+1)w_{+}
+[(2m-1)\delta-1]w_{-}+m\delta=0.
\label{cf.mdelta}
\end{equation}

In this case, the solution in analytic form is available:
We obtain explicitly
\begin{eqnarray}
w_+(T) & = & \frac{1}{2(\delta +1)}\biggl\{-m\delta [f(T)+1]
\pm \sqrt{m^2 \delta^2[f(T)+1]^2-4(\delta+1)[(2m-1) \delta
-1] f(T)}\biggr\}, \nonumber \\
w_-(T) & = & \frac {f(T)}{w_+(T)}~.
\label{cf.wpm}
\end{eqnarray}
The upper sign is for $\delta>0$ and the lower sign
for $\delta<0$. At $T=0$, eq. (\ref{cf.wpm}) indeed yields
$n_+ = m\delta, n_- = 0$,~for $\delta>0$, and similarly
$n_+ =0,  n_- = -m\delta$, for $\delta<0$, as expected from
eq. (\ref{n+-n-}).

Numerical results for a two-dimensional plot of
$n_+(T,\delta)$ and $n_-(T,\delta)$ for filling
factors near $1/m=1/3$ are shown
in Figures \ref{fig-1-nplus} (b) and
\ref{fig-2-nminus} (b).

\subsection{Correlated Vortex or Projected Composite Fermion Picture}

{}For the statistics matrix (\ref{alpha-vortex2}),  eq.
(\ref{npm}) is explicitly

\begin{equation}
n_{+}={ w_{-}+2\over (w_{+}+2-1/m) (w_{-}+1/m) +(2-1/m)^2 }~,
\label{cv.n_+}
\end{equation}
and
\begin{equation}
n_{-}= { w_{+}\over (w_{+}+2-1/m) (w_{-}+1/m) +(2-1/m)^2 }~,
\label{cv.n_-}
\end{equation}
with $w_+$ and $w_-$ satisfying
\begin{equation}
m \delta w_+ w_{-} +(\delta+1)w_{+}
+[(2m-1)\delta-1]w_{-}+2(2m-1)\delta-2 =0,
\label{cv.mdelta}
\end{equation}
\begin{equation}
w_{-}^2(1+w_+)(1+w_-)^{-1}=f(T).
\label{cv.f(T)}
\end{equation}

Again, analytic solution is impossible, but
a numerical two-dimensional plot  for
$n_+(T,\delta)$ and $n_-(T,\delta)$ is shown,
respectively,  in Figures \ref{fig-1-nplus} (c)
and \ref{fig-2-nminus} (c), for filling factors
near $1/m=1/3$.

\subsection{Low-Temperature Asymptotics}

As application of the above explicit formulas,
let us discuss the low-temperature asymptotics
of  the density of activated pairs. For simplicity,
we consider the case with exactly $\nu=1/m$, or
$\delta=0$. It is easy to check that in either
of the schemes, we have $n_{+}(T)=n_{-}(T)$.
At very low temperatures, $f(T)$ is very large,
so we have $w_{\pm}\approx \exp \{\Delta / 2kT\}$.
This leads to

\begin{equation}
\rho_{\pm} (T) \approx \rho_{0}\exp \{\Delta / 2kT\}\, ,
\label{asymp}
\end{equation}

\noindent with the prefactor $\rho_0 \equiv
G_{\pm}/V = (1/m) N_{\phi}/V$, proportional
to the (fractional) quasiparticle charge.

This is in complete agreement with the standard
Boltzmann behavior characteristic of thermal
activation across a finite gap. Note that this
behavior is independent of the statistics matrix.
Thus to look for the effects of fractional
exclusion statistics, the temperature should
be higher than this asymptotic region.

\section {Emergence of Hierarchical States}

{}From the general equation of state (\ref{Omega}),
one can easily see that the pressure $P$ will diverges,
when one of the denominators in the right side
becomes zero, i.e. when the excitation densities satisfy

\begin{equation}
\sum_{j=+,-} \alpha_{ij} n_{j}(T)=1~~~
(i = +  ~{\rm or}~ -).
\label{newstate1}
\end{equation}
This corresponds to the situation in which one of the
$G_{eff,\mp}^{(b)}$ vanishes (see  eq. (\ref{Geff})),
so that there is no available quasihole or quasielectron
states for additional thermally activated pair to occupy.
Using eq.  (\ref{n(T)}), the condition (\ref{newstate2})
is reduced to

\begin{equation}
w_{+}n_{+}=0,~~~~~
{\rm or} ~~~~  w_{-}n_{-}=0
\label{newstate2}.
\end{equation}

At zero temperature in the ground state with the
filling factor $\nu$ near the magic $\nu_0=1/m$
($m$ odd), there is no thermally activated pair,
so one of the $n_{\pm}$ vanishes, depending on
whether $\nu$ is greater or smaller than $1/m$.
There are three cases that the condition
(\ref{newstate2}) is satisfied:

\noindent (i) $n_+ = n_- =0$: This is the case
with $\nu=\nu_0$, which is just the original Laughlin's
$1/m$ incompressible state.


\noindent (ii) $n_-=0$ and $n_+ >0$: According
to the relation (\ref{n+-n-}), this is the case with
$\nu> \nu_0$.  Then $w_+=0$, and eq. (\ref{wkt})
requires $w_-=\infty$. So from eq. (\ref{npm}), one
obtains $n_+ = 1/\alpha_{++}$. For
$\alpha_{++}=2-1/m$, it leads to $\nu=2/(2m-1)$.
This gives rise to a new incompressible state in
which quasiholes are absent and quasielectrons
fill up all possible states. In the bosonic vortex
scheme, this new quantum Hall state is called the first
hierarchical state \cite{Hald1,Halperin.hier},
in which the hard-core quasielectrons in the
Laughlin $1/m$-state are condensed to form a
new incompressible liquid. If quasielectrons did not
have hard core, they would not be able to form
the new incompressible hierarchical state.
In the composite fermion scheme, the original
Laughlin $1/m$-state corresponds to $n=1$
in eq. (\ref{filling}), and is interpreted as
complete filling of the lowest Landau level in the
residue magnetic field, while the new state
corresponds to $n=2$ with the second Landau
level completely filled by composite fermions.

\noindent (iii) $n_+=0$ and $n_- > 0$: According
to the relation (\ref{n+-n-}), this is the case with
$\nu< \nu_0$.  Then $w_-=0$, and eq. (\ref{wkt})
requires $w_+ =\infty$. So from eq. (\ref{npm}), one
obtains $n_- = 1/\alpha_{--}$. For $\alpha_{--}=1/m$,
it leads to $\nu=1-1/m$ state, in conflict with the
condition $\nu< \nu_0$. So unlike quasielectrons,
the quasiholes in the Laughlin $1/m$-liquid can not
condense to form a new incompressible state.
Numerical data presented in ref. \cite{16} confirms
this conclusion.

We note that in the above discussions for $T=0$,
mutual statistics is irrelevant.

What will happen at finite $T$? Suppose that
$\nu > \nu_0$. Due to  thermal activation of
quasiparticle pairs, both $n_{\pm}(T)>0$,
so there is an additional contribution in
eq. (\ref{newstate1}) from thermally activated
quasiholes, which is $T$-dependent. One may
wonder if the effect of mutual statistics would
lead to formation of new incompressible
(hierarchical) states at $T$-dependent filling
factors. At finite $T$, the condition
(\ref{newstate2}) is reduced to $w_{+} (T) =0$
or $w_{-}(T)=0$. It can be shown, case by case
for the three statistics matrices (\ref{alpha-vortex2}),
(\ref{jainstat}) and (\ref{jainstat2}), that in either
scheme, {\it the pressure can not diverge at any
filling factor}. For example, in the composite
fermion scheme, eq. (\ref{f(T)}) with finite $T$
implies that  if $w_{+} (T) =0$, then
$w_{-}(T)=\infty$, which in turn leads to
$n_{-}(T)=0$, in accordance to eq. (\ref{cf.n_-}).
But this is impossible at finite $T$ due to thermal
activation. Similarly  $w_{-} (T) =0$ would
lead to $n_{+}(T)=0$, a contradiction too.
The same is true in the other two schemes.

The fact that the pressure $P$ never diverges
at finite $T$ means that the quantum Hall
transitions due to a divergent $P$ are a quantum
phase transition at zero temperature.

\section{Thermodynamic Observables}

Under the assumption that quasiparticles
dominate the low-lying excitation spectrum
of the FQH liquids, the knowledge of thermal
activation of quasiparticle pairs allows
us to calculate thermodynamic observables
of the FQH liquids. In this section, we will
show numerical results with three possible
statistics matrices (\ref{alpha-vortex2})
(\ref{jainstat}) and (\ref{jainstat2}).

In Figure \ref{fig-3-pv}, we show the
temperature dependence of the thermonynamic potential $pV$ for several
different filling factors.

The average energy density $h(T)$ of the
quasielectron-quasihole pairs is given as
\begin{eqnarray}\label{eq:dgam-dt3}
h(T) =\varepsilon_{+} [(\rho_+(T)-\rho_+(0)]
+\varepsilon_-[(\rho_-(T)-\rho_-(0)].
\end{eqnarray}
 
The specific heat $C_v (T, \delta)$ is also
evaluated numerically using the expression
\begin{equation}
C_v=\frac{\partial h(T) }{\partial T}=-k\rho_0
E^2\frac{\partial n_-}{\partial E},
\end{equation}
where $E= \Delta/k T$.
In Figure \ref{fig-4-spc}, we have shown
the curves of $C_v(T)$ at different values of
$\delta$ for the three statistics matrix.
At very low temperatures, $C_v$ vanishes
exponentially: $C_v(T)\sim \exp\{-\Delta/kT\}$,
due to a finite activation gap. At higher
temperatures, $C_v$ increases due to quasiparticle
pair creation. When the temperature approaches the
order of magnitude of the gap energy $\Delta$, the
difference between different mutual statistics
becomes more and more apparent. The phenomenon
that $C_v (T)$ decreases after reaching a maximum
is a manifestation of exclusion statistics due to
the saturation of the available states for
quasiparticles. The position of the maximum of
the $C_v (T)$ curve is statistics-matrix dependent.
Whether our approximation is still valid at this
temperature or not is a question we cannot answer
in our approach. We leave it to experiments.
However, we are sure that the vanishing of $C_v$
at very high temperatures should not be trusted,
since our approximation of restricting quasielectrons
to the lowest available Landau level certainly
breaks down.

Now let us consider the magnetic response of the system.
The magnetization per unit area due to quasiparticle
excitations has been given in ref. \cite{3} as
\begin{eqnarray}
{\cal M}&=&-\biggl(\frac{\partial \Omega/V}
{\partial B}\biggr)_{T,V,\mu_i/kT}\nonumber \\[3mm]
&=&\displaystyle\sum_{i=+,-}
\biggl(-\mu^B_i\rho_i+\frac{kT}{3\phi_0}
\ln\frac{1+w_i}{w_i}\biggr).
\end{eqnarray}
Here $\phi_0=hc/e$. In the derivation, we have
effectively defined $\mu_\pm^B =\partial \varepsilon_\pm
/ \partial B$. In comparison with experimental data,
the effective magnetic moment, $\mu_\pm^B$, of a single
quasiparticle could be either treated as phenomenological
parameters or derived from some microscopic model.
In the following, we will assume $\mu_\pm$ to be
$B$-independent.

This equation can be viewed as a generalization
of the case with one species of anyons in the
lowest Landau level. \cite{Das1,3}
The first term, corresponding to the usual
de Haas-van Alphen term, is of the same form
in the two cases. However, there are important
differences between the one-species anyon case
and the present situation: For the former case,
the density of anyons, $\rho$, is fixed either
when $B$ or $T$ varies (no thermal activation
of anyon pairs). The resulting susceptibility
has a simple analytic expression and
vanishes as $T \to 0$.\cite{3} However,
this is not true in the present case, where the
densities of quasielectrons and quasiholes
are both $B$- and $T$-dependent.
Even at zero temperature, the magnetization has 
a $B$-dependence through that of $\rho_{\pm}$:
\begin{equation}
{\cal M}(T=0)=\mu^B_{\pm}\rho_{\pm}(B,T=0)
=\mu_{\pm}^B \rho_0 n_{\pm}(B,T=0).
\end{equation}
(The subscript $\pm$ depends on whether $\nu-1/m$
is positive or negative.) This implies that the
zero-temperature susceptibility is non-zero, in
contrast to the case of anyons of one species:
\begin{equation}
\chi(T=0)=\biggl(\frac{\partial{\cal M}(T=0)}
{\partial B}\biggr)_V=\frac{\mu_+^B}{\phi_0}.
\end{equation}

Our general expression for the finite-temperature
susceptibility reads
\begin{equation}
\chi=\biggl(\frac{\partial{\cal M}}
{\partial B}\biggr)_{T,V}=\chi_0+\chi_1,
\end{equation}
where $\chi_0$ is a de Haas-van Alphin-like term and
$\chi_1$ comes from thermally activated pairs.
They are given by
\begin{eqnarray}
\chi_0&=&\frac{\mu_+^B}{3\phi_0}\sum_{i=+,-}
\biggl(-n_i+(\delta+1)\frac{\partial n_i}
{\partial \delta}\biggr),\nonumber \\[3mm]
\chi_1&=&\frac{\Delta}{9\rho_e\phi_0^2}\sum_{i=+,-}
\frac{(\delta+1)^2}{E w_i(1+w_i)}
\frac{\partial w_i}{\partial \delta},
\end{eqnarray}
where we have assumed $\mu_+^B =\mu_-^B$
for simplicity.
Figures \ref{fig-5-chi0} and \ref{fig-6-chi1} show
the susceptibilities $\chi_0$ and $\chi_1$ as
functions of $T$ and $\delta=3\nu-1$ (with
$0 \leq \delta \leq 1/5$ corresponding to
$1/3 \leq \nu \leq 2/5$), for three different
counting schemes for statistics matrix. Indeed, for 
$1/3<\nu<2/5$, $\chi_1\propto T$ and tends to zero
as $T\to 0$, so that $\chi\to\chi_0(T=0)$ as
expected. When $\Delta\sim 100$ mK,
$\rho_e\sim 10^{11}$ cm$^{-2}$ and $\mu_+^B
\sim \mu_0^B$ (Bohr's magnetic moment),
the ratio of the coefficients of $\chi_0$, 
$A_0=\mu_+^B/3\phi_0$, and that of
$\chi_1$, $A_1=\Delta/9\rho_e\phi_0^2$,
is $A_0/A_1 \sim 10^2$. Therefore, in a certain range
of filling factor greater than $\nu_0=1/3$, the
de Haas-van Alphin term dominates, which is positive
at $T=0$. On the other hand, from Figure
\ref{fig-5-chi0}, we see that $\chi_1$ is always
negative in the range of fillings at hand. As $T$
increases, the magnitude of $\chi_1$ increases,
gradually becomes comparable to $\chi_0$ and 
eventually dominates. Then the total susceptibility
will change sign. Thus, we observe that for a given
filling factor, there can be a paramagnetism-diamagnetism
transition, with the critical temperature generically
$T_c \sim \Delta$. The new incompressible state with
$\nu=2/5$ at $T=0$ is totally diamagnetic, because 
with $\nu \to 2/5$ one has $w_\to 0$, which causes 
$\chi_1$ divergent. This might be viewed as a 
generalization of Landau diamagnetism to quasielectrons.

\section{Conclusions and Discussions}

Quasielectrons and quasiholes in the FQH
liquids obey fractional (including nontrivial
mutual) exclusion statistics. Their statistics matrix
near the magic electron filling $1/m$ (with $m$ odd)
can be determined from various state-counting
schemes. These schemes involve different
assumptions on statistical correlations between
quasiparticles, resulting in somewhat different
mutual statistics. A common feature of the
schemes is that both quasiholes and quasielectrons
in the incompressible FQH liquid background
behave like a charge in an effective magnetic field.
The common assumption is that quasielectrons
are in the lowest available Landau level
(with respect to the effective magnetic field).
Then the thermal activation of FQH
quasiparticles at low temperatures is
discussed in the dilute generalized ideal gas
approximation. If these quasiparticles dominate
the low-lying excitation spectrum, their
contributions dominate the low-temperature
thermodynamics. Otherwise, contributions from
other low-lying elementary excitations (such as
skyrmions, if they exist and are important) have to
be added to the quasiparticle contributions we
obtain here. At higher temperatures, our
assumptions obviously break down. We hope the
situation could be improved in the future by
incorporating corrections. Right now we just
leave the question of when the corrections
should be important open to experiments.

We have used the approximation of generalized
ideal gas in our treatment of statistical
thermodynamics. It is good only when the
thermally activated quasiparticle pairs are not
too dense. On the other hand, to
look for the effects of mutual statistics, the
densities of quasiparticle pairs should not be
too low. We hope there is some intermediate
range for quasiparticle densities in which the
two conflicting requirements could be
reconciled to some extent. Whether this is true,
only experiments can tell.

It is shown that the existence of hierarchical
states in the FQH effect can be viewed as a
manifestation of the exclusonic nature of
the relevant quasiparticles. The associated
FQH phase transition is shown to occur only at
zero temperature. Thermal activation of
quasiparticle pairs and thermodynamic
observables are numerically studied with three
possible statistical matrices. At zero temperature,
they are all equivalent to each other, but
differences show up at finite temperature.
In particular, we have demonstrated that for
a fixed filling factor between $1/3$ and $2/5$,
with increasing temperature, the system
may possibly exhibit a transition from
paramagnetism to diamagnetism. However,
we should be cautious about this possibility:
We have assumed that $\mu_{\pm}^B$ are of
the same order of magnitude as the Bohr magneton;
also  the approximation of generalized ideal gas
might break down at the would-be transition
temperature.

It is desirable that these theoretical
predictions would be put to experimental tests,
if the tremendous difficulties in measuring
thermodynamic quantities of a thin layer of
electron gas could be overcome someday.

\section{Acknowledgments}

We acknowledges helpful discussions with Jian Yang and
Fu-Chun Zhang. The work of Y.S.W. was supported in 
part by U.S. NSF grant PHY-9601277, that of Y. Y.
by NSF of China, and that of Y.H. by Grant-in-Aid
from the Ministry of Education, Science and Culture
of Japan. The computation in this work has been partly 
done using the facilities of the Supercomputer Center, 
ISSP, University of Tokyo.

\newcommand{\bibit}{\it}
\newcommand{\bibbf}{\bf}
\renewenvironment{thebibliography}[1]
        {\begin{list}{\arabic{enumi}.}
        {\usecounter{enumi}\setlength{\parsep}{0pt}
\setlength{\leftmargin 1.25cm}{\rightmargin 0pt}
         \setlength{\itemsep}{0pt} \settowidth
        {\labelwidth}{#1.}\sloppy}}{\end{list}}


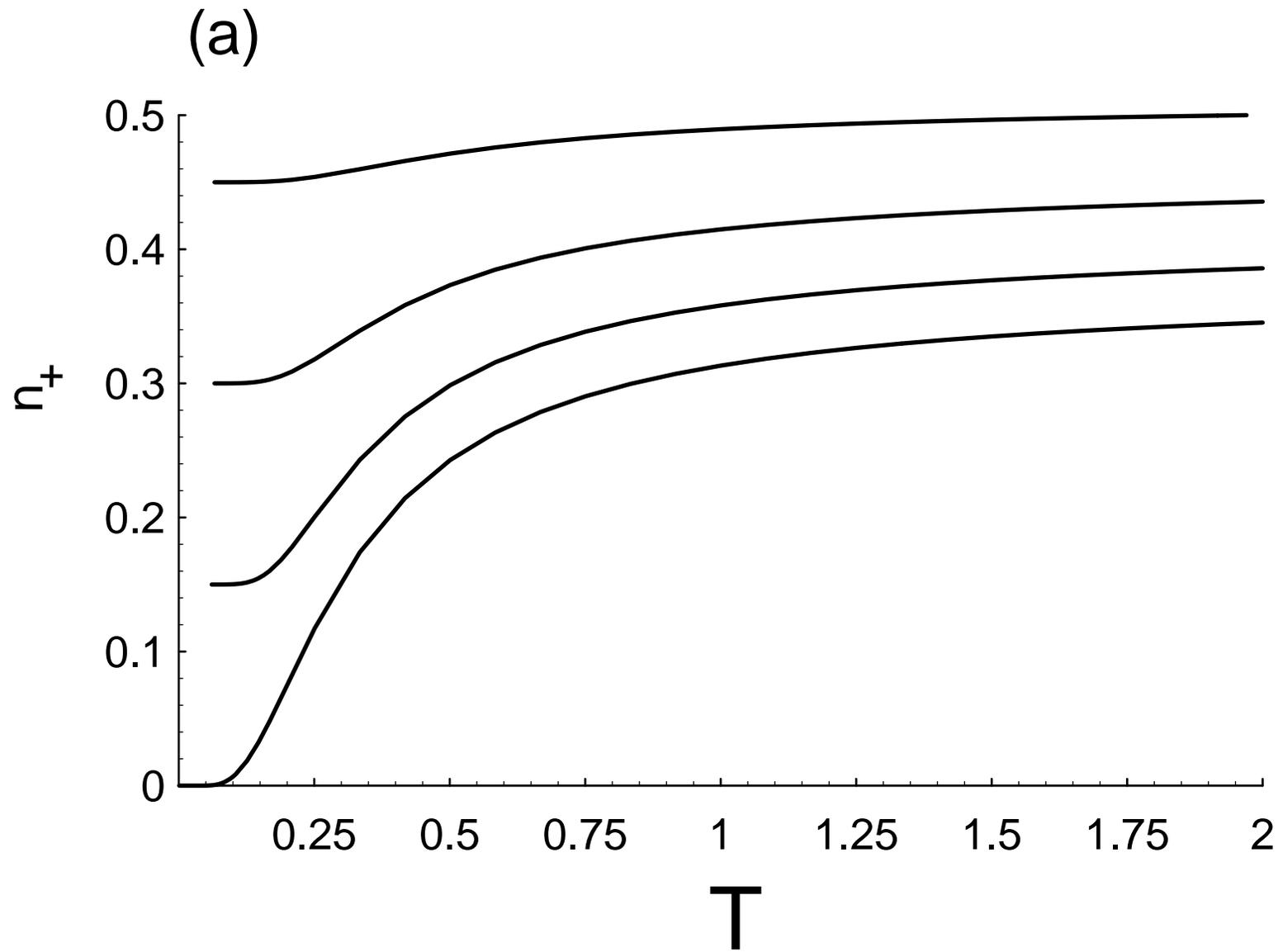
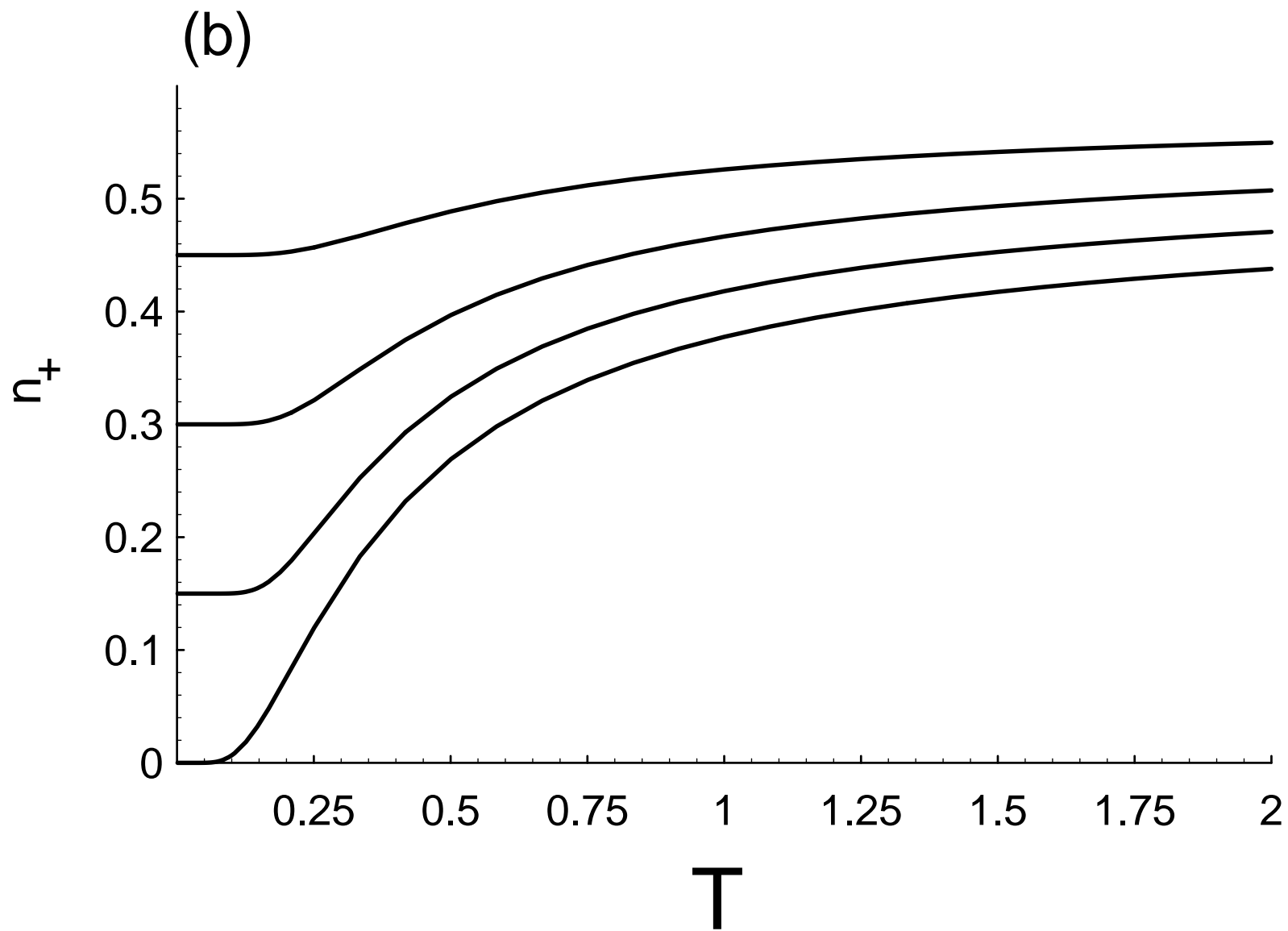
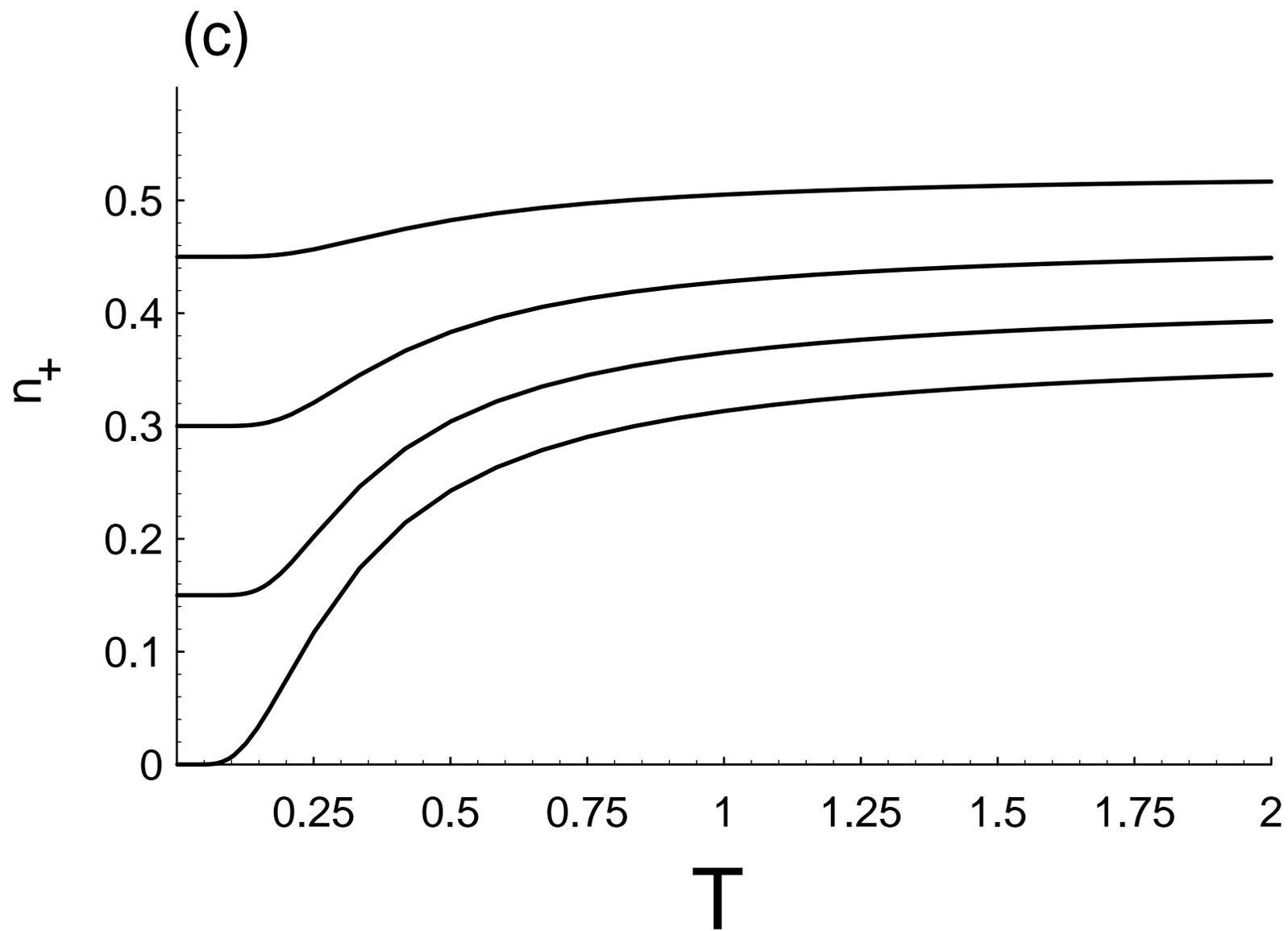
\begin{figure}
\caption{
The two dimensional plot of the occupation number of
the quasielectron $n_+(T,\delta)$.
The filling factor is near $1/m=1/3$.
(a) for the bosonic vortex scheme,
(b) for the composite fermion scheme,
(c) for the correlated vortex or projected composite fermion scheme.
In all the figures, $\delta=0.15$, $0.1$, $0.05$, and $0$ respectively,
 from above.
 \label{fig-1-nplus}}
\end{figure}

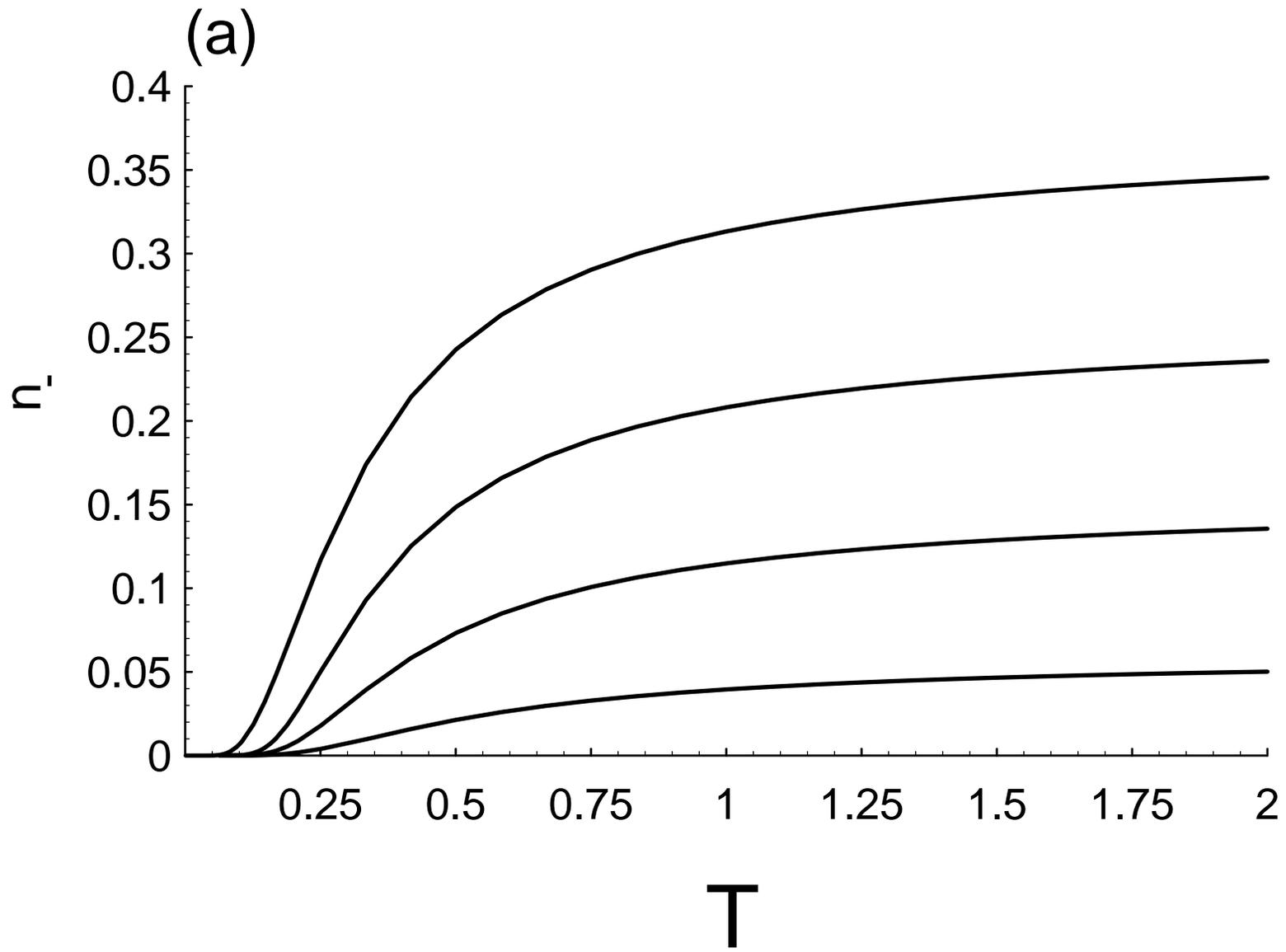
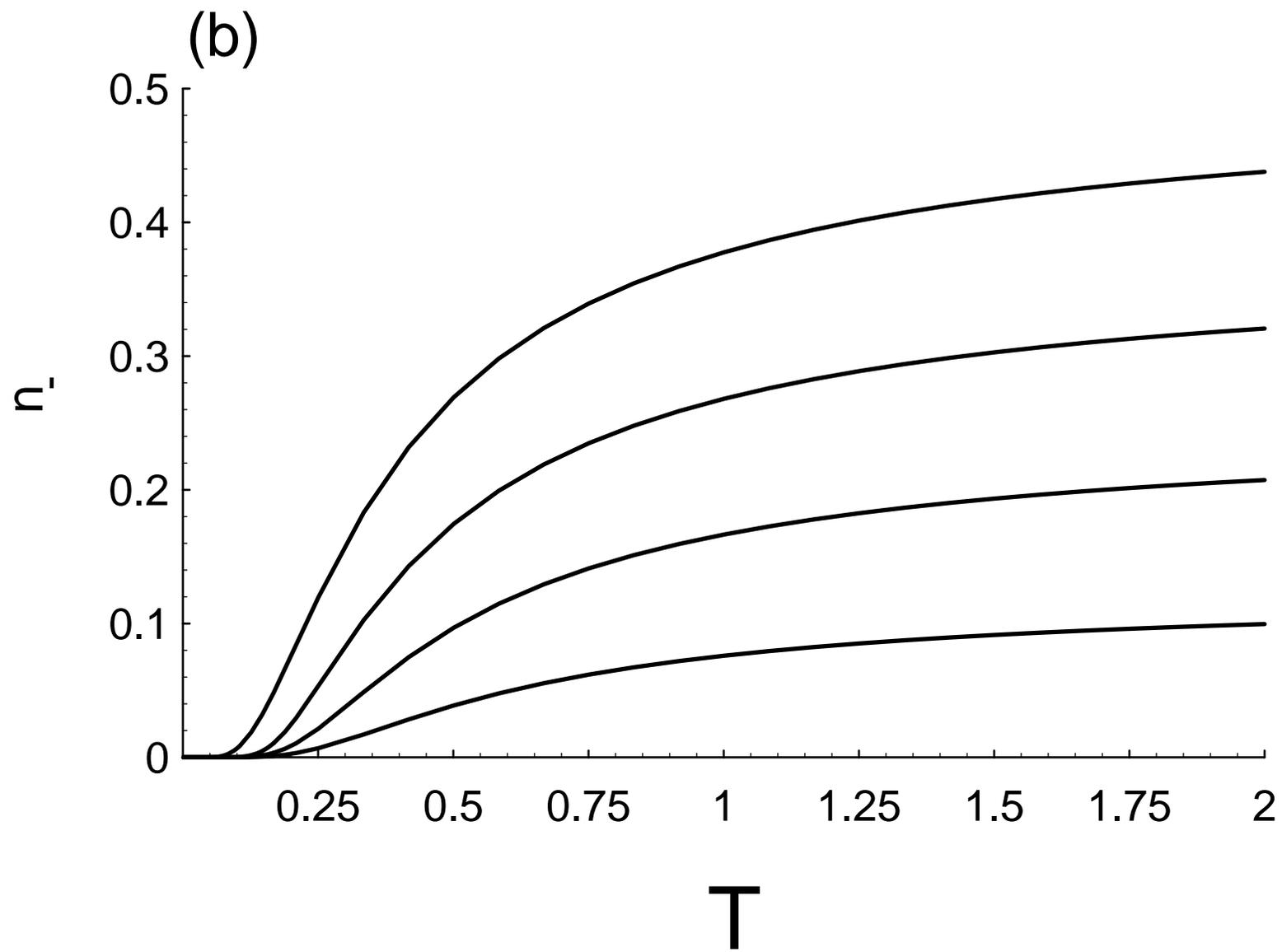
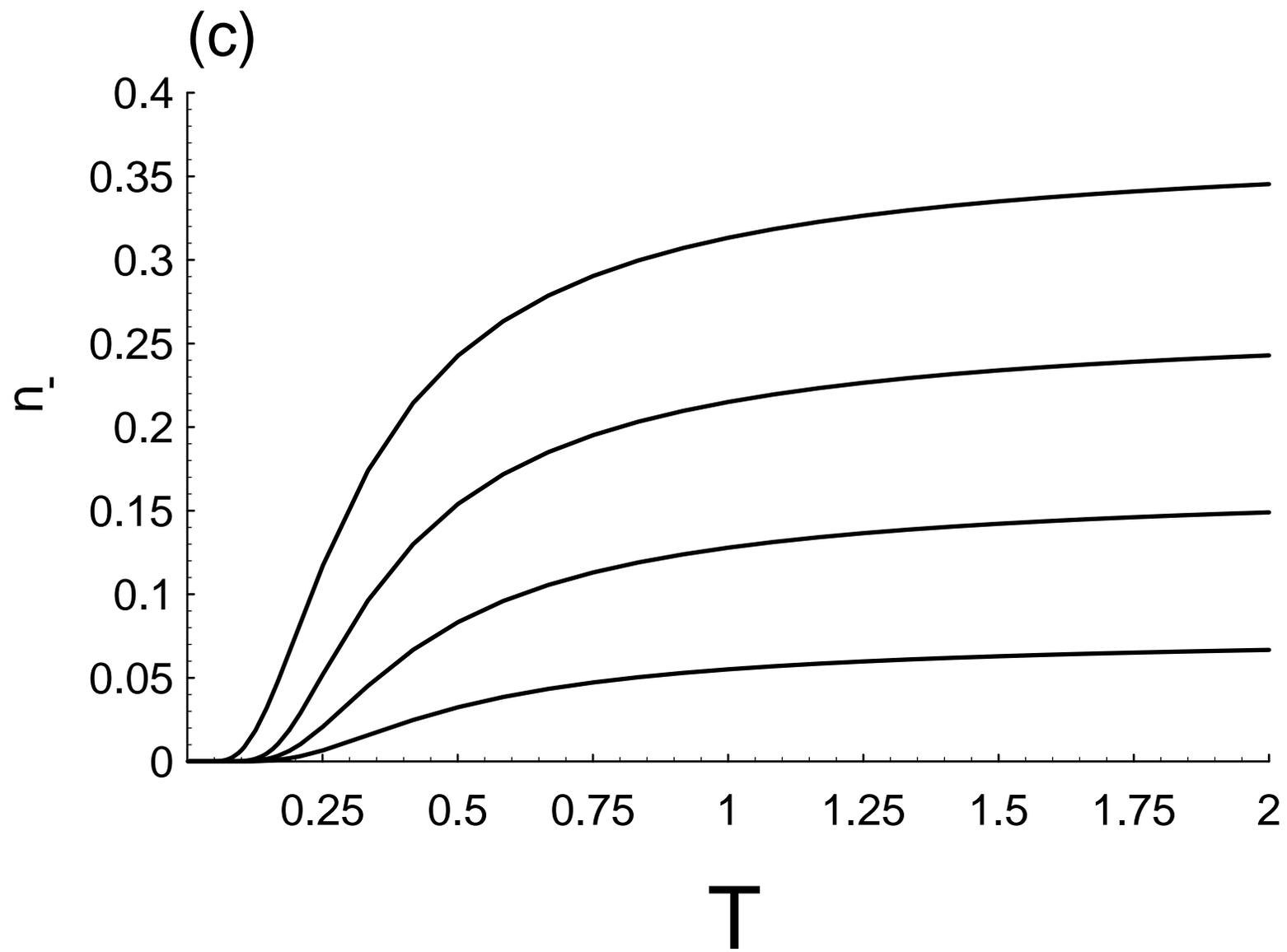
\begin{figure}
\caption{
The two dimensional plot of the occupation number of
 the quasihole $n_-(T,\delta)$.
The filling factor is near $1/m=1/3$.
(a) for the bosonic vortex scheme,
(b) for the composite fermion scheme,
(c) for the correlated vortex or projected composite fermion scheme.
In all the figures, $\delta=0$, $0.05$, $0.1$, and $0.15$ respectively,
 from above.
\label{fig-2-nminus}}
\end{figure}

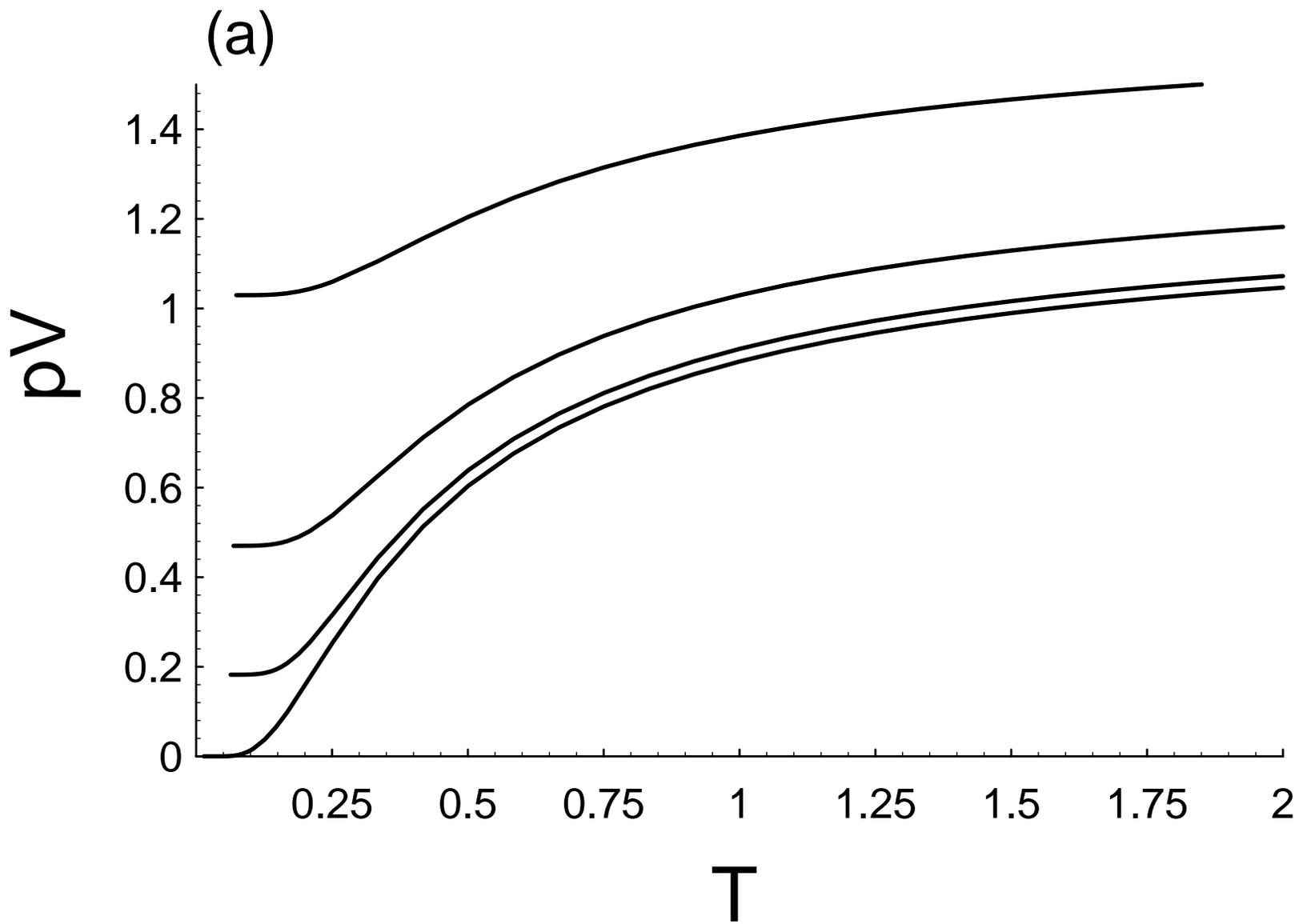
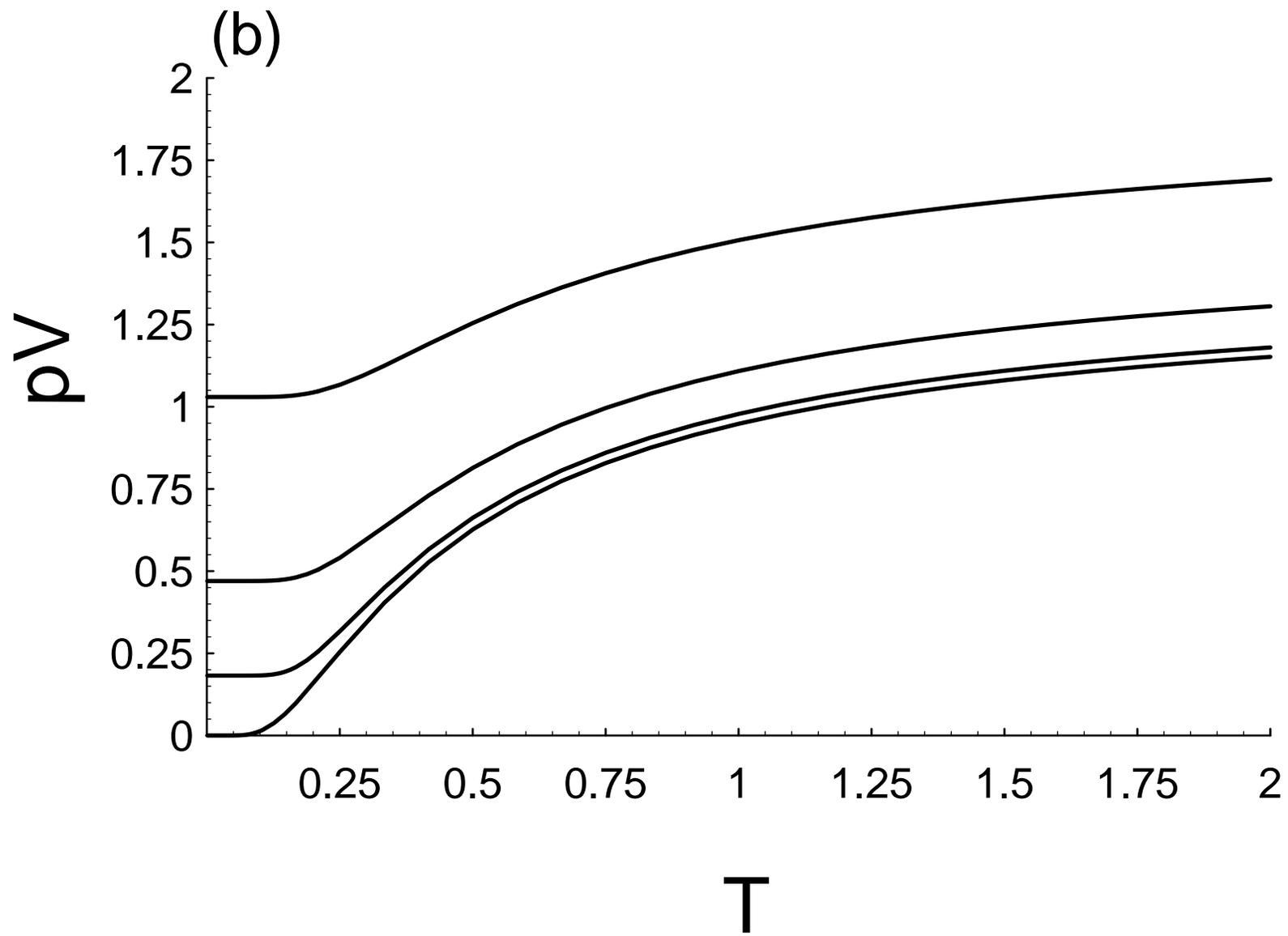
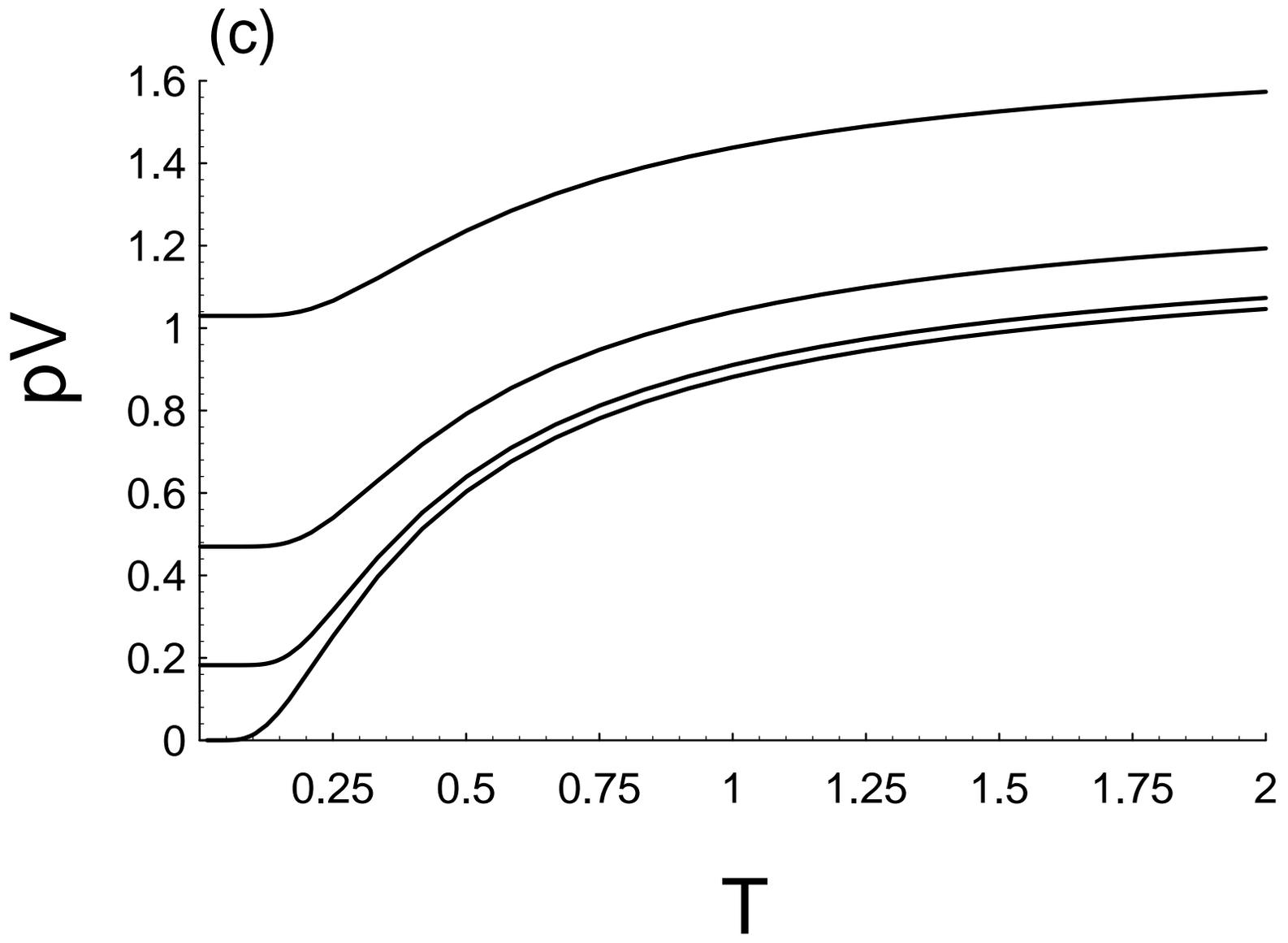
\begin{figure}
\caption{
The thermodynamic potential $\Omega = pV$   near
the filling factor  $ 1/3$.
(a) for the bosonic vortex scheme,
(b) for the composite fermion scheme,
(c) for the correlated vortex or projected composite fermion scheme.
In all the figures, $\delta=0.15$, $0.1$, $0.05$, and $0$ respectively,
 from above.
\label{fig-3-pv}}
\end{figure}

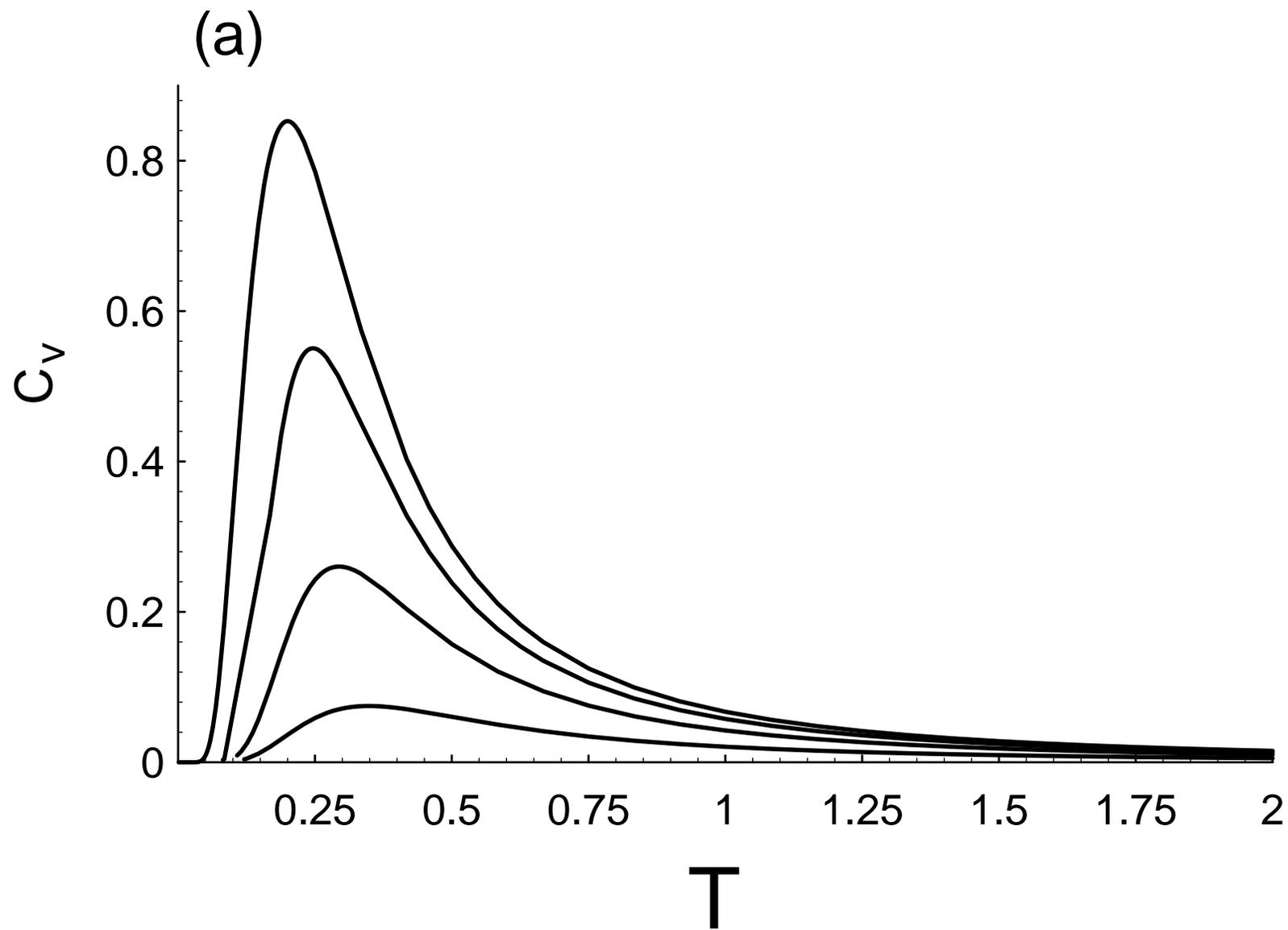
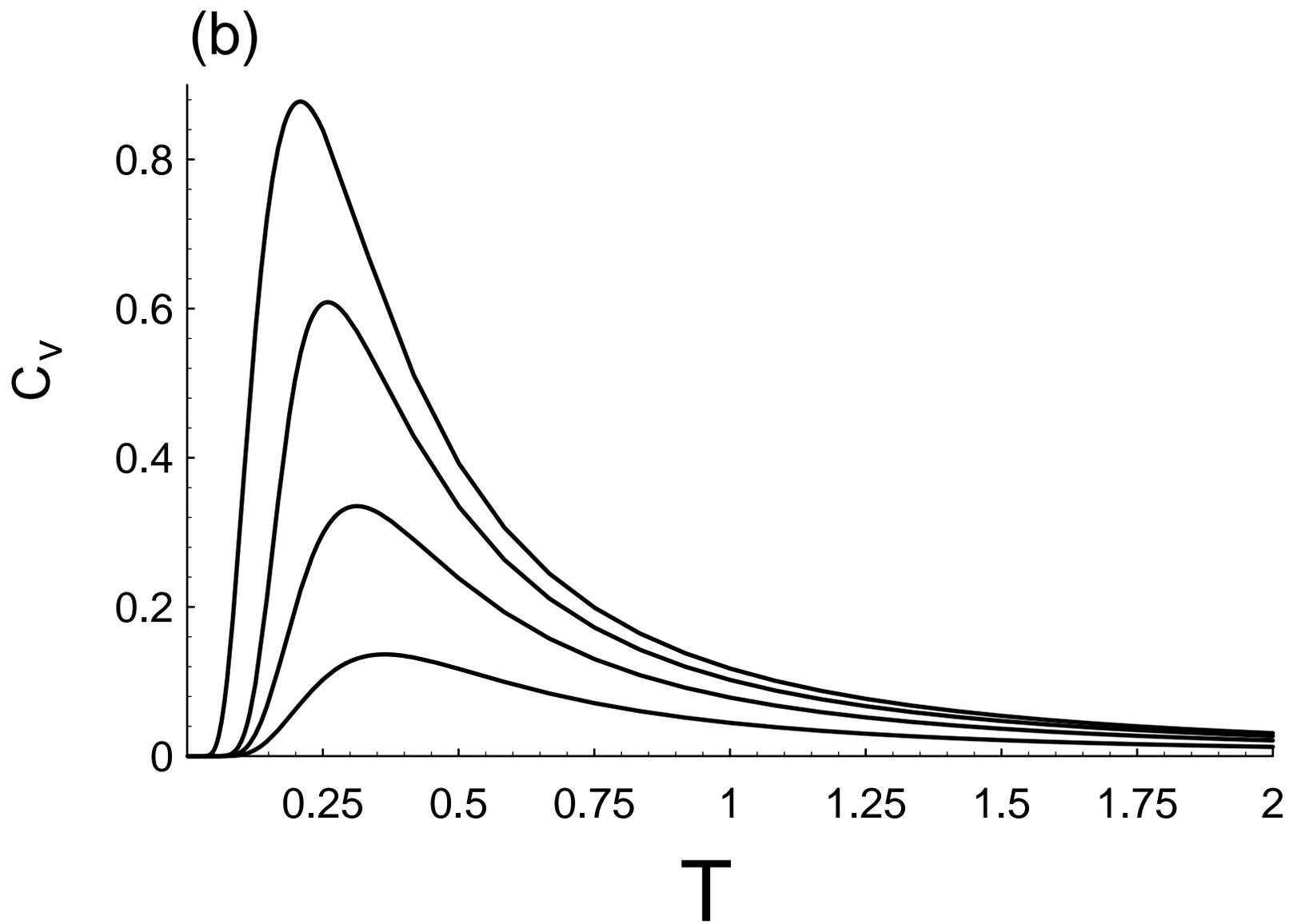
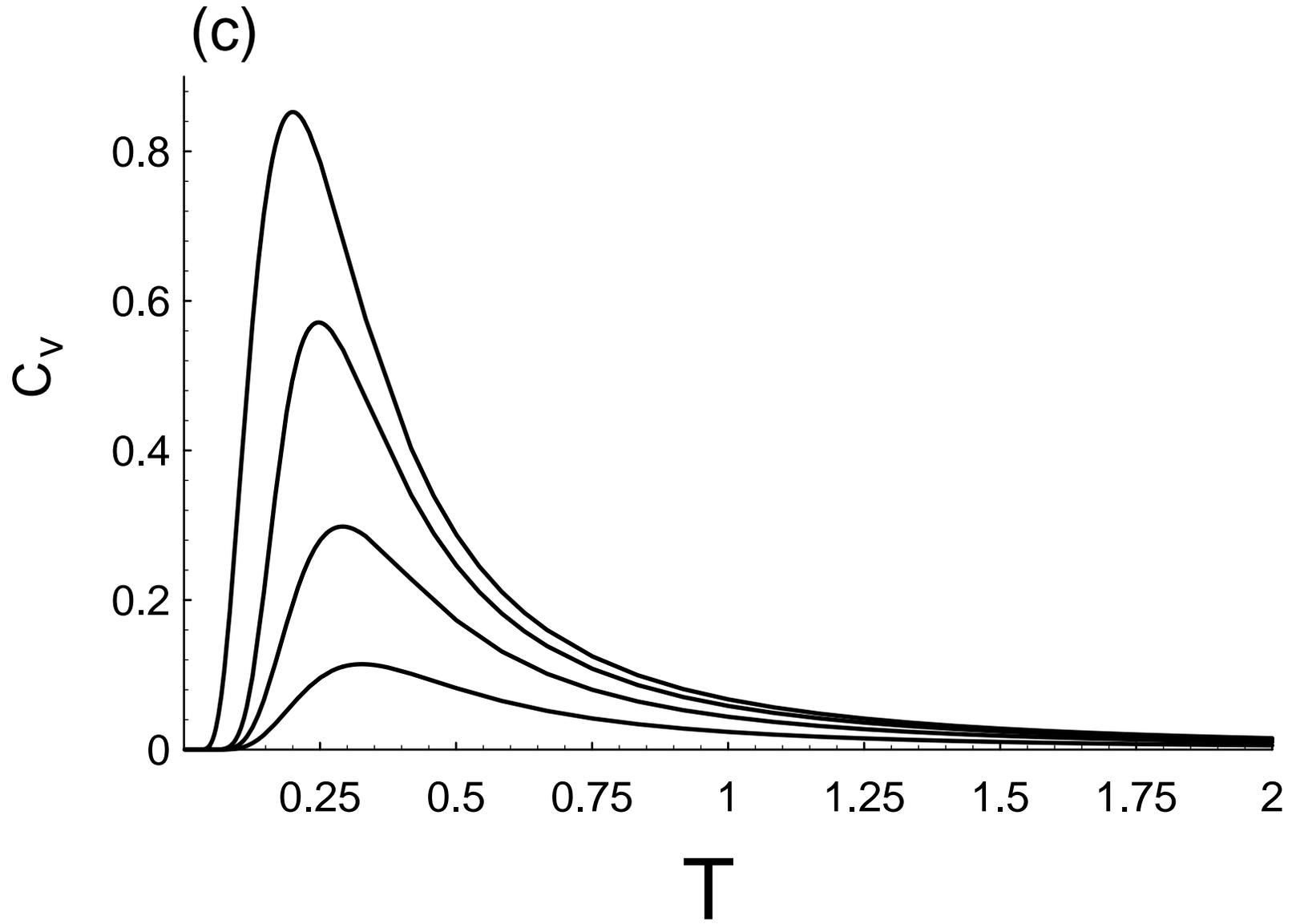
\begin{figure}
\caption{
The temperature dependence of the specific heat
$C_v(T,\delta) $  near
the filling factor  $ 1/3$.
The different curves are different $\delta$'s.
(From above, $\delta=0,\ 0.05,\ 0.1,\ 0.15$.)
(a) for the bosonic vortex scheme,
(b) for the composite fermion scheme,
(c) for the correlated vortex or projected composite fermion scheme.
In all the figures, $\delta=0$, $0.05$, $0.1$, and $0.15$ respectively,
 from above.
\label{fig-4-spc}}
\end{figure}

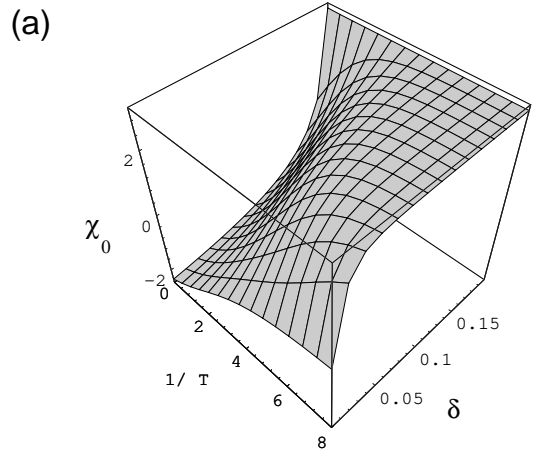
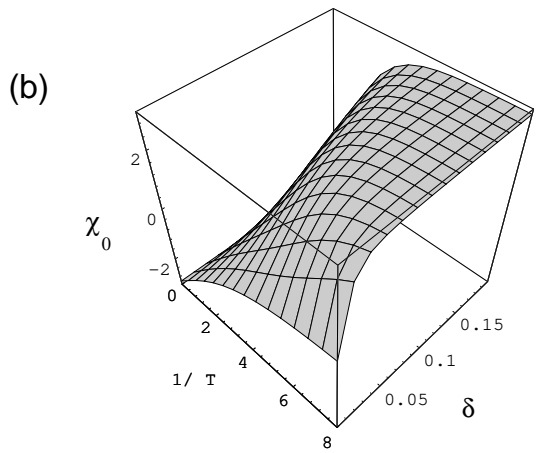
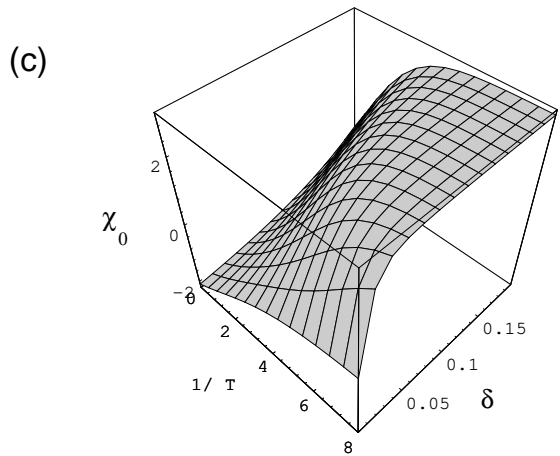
\begin{figure}
\caption{
The two dimensional plot of
$\chi_0(T,\delta) $ (de Haas-van Alphin term of the susceptibility ) near
the filling factor  $ 1/3$.
(a) for the bosonic vortex scheme,
(b) for the composite fermion scheme,
(c) for the correlated vortex or projected composite fermion scheme.
\label{fig-5-chi0}}
\end{figure}

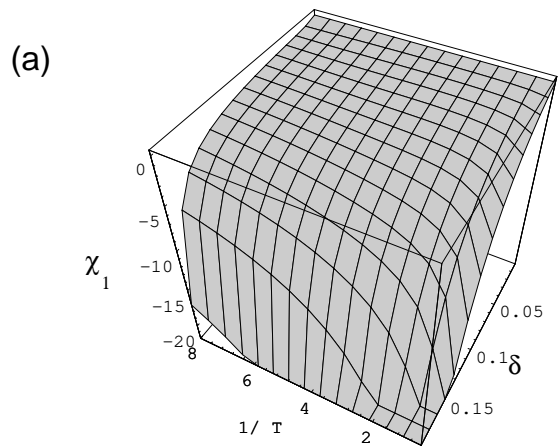
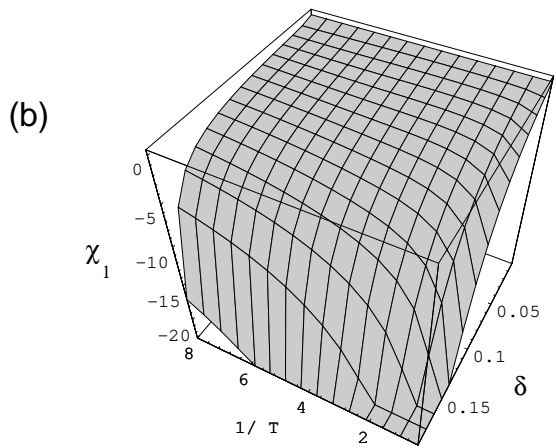
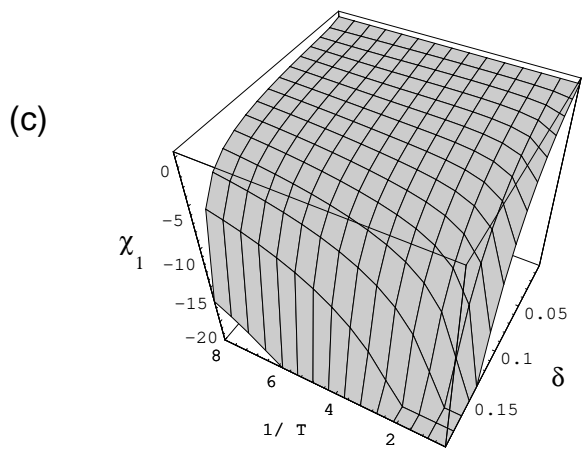
\begin{figure}
\caption{
The two dimensional plot of$\chi_1(T,\delta) $ ( a pair excitation term of
the susceptibility ) near
the filling factor  $ 1/3$.
(a) for the bosonic vortex scheme,
(b) for the composite fermion scheme,
(c) for the correlated vortex or projected composite fermion scheme.
\label{fig-6-chi1}}
\end{figure}



\begin{references}
\bibitem{comm1}
Here we have ignored a constant term of order of
unity in the right side of the equation, which is
unimportant in the thermodynamic limit. Similar
situation may happen later.
\end{references}
\end{document}